\documentclass{emulateapj}
\usepackage{amsmath}
\usepackage{mathtools}
\usepackage{epstopdf}
\usepackage{natbib}

\newcommand{\emcee}{\texttt{emcee}}

\def\spose#1{\hbox to 0pt{#1\hss}}
\def\simlt{\mathrel{\spose{\lower 3pt\hbox{$\mathchar"218$}}
     \raise 2.0pt\hbox{$\mathchar"13C$}}}
\def\simgt{\mathrel{\spose{\lower 3pt\hbox{$\mathchar"218$}}
     \raise 2.0pt\hbox{$\mathchar"13E$}}}

\shorttitle{M31 Structural Decomposition}
\shortauthors{Dorman et al.}
\slugcomment{Accepted for publication in the Astrophysical Journal}
\begin{document}
\bibliographystyle{plainnat}

\title{
A new approach to detailed structural decomposition \\
 from the SPLASH and PHAT surveys: \\
Kicked-up disk stars in the Andromeda Galaxy? \\
} 
\author{Claire~E.\ Dorman\altaffilmark{1}, 
Lawrence~M.\ Widrow\altaffilmark{2},
Puragra\ Guhathakurta\altaffilmark{1}, 
Anil~C.\ Seth\altaffilmark{3},
Daniel\ Foreman-Mackey\altaffilmark{4}, 
Eric~F.\ Bell\altaffilmark{5},
Julianne~J.\ Dalcanton\altaffilmark{6},
Karoline~M.\ Gilbert\altaffilmark{6, 7, 8},
Evan~D.\ Skillman\altaffilmark{9},
Benjamin~F.\ Williams\altaffilmark{6}
}

\altaffiltext{1}{UCO/Lick Observatory, University of California
    at Santa Cruz, 1156 High Street, Santa Cruz, CA~95064, USA; {\tt [cdorman,
      raja]@ucolick.org}}
\altaffiltext{2}{Department of Physics, Engineering Physics, and
  Astronomy, Queen's University, Kingston, Ontario, Canada; {\tt
    widrow@astro.queensu.ca}}
\altaffiltext{3}{Department of Physics \& Astronomy, University of
  Utah, Salt Lake City, UT 84112, USA; {\tt aseth@astro.utah.edu}}
\altaffiltext{4}{Department of Physics, New York University, New York,
  NY~10003, USA; {\tt danfm@nyu.edu}}
\altaffiltext{5}{Department of Astronomy, University of MIchigan, 500
  Church Street, Ann Arbor, MI~48109, USA; {\tt ericbell@umich.edu}}
\altaffiltext{6}{Department of Astronomy, University of Washington,
  Box 351580, Seattle, WA 98195; {\tt [jd, kgilbert, ben]@astro.washington.edu}}
\altaffiltext{7}{Hubble Fellow}
\altaffiltext{8}{Current address: Space Telescope Science Institute, 
3700 San Martin Drive, Baltimore, MD 21218, USA}
\altaffiltext{9}{Minnesota Institute for Astrophysics, School of
  Physics and Astronomy, University of Minnesota, 116 Church St. SE,
  Minneapolis, MN 55455, USA; {\tt skillman@astro.umn.edu}}
\begin{abstract}

We characterize the bulge, disk, and halo subcomponents in the
Andromeda galaxy (M31) over the radial range $4~{\rm kpc}<R_{\rm
  proj}<225$ kpc. The cospatial nature of these subcomponents renders
them difficult to disentangle using surface brightness (SB)
information alone, especially interior to $\sim 20$ kpc. 
Our new decomposition technique combines information
from the luminosity function (LF) of over $1.5$ million bright $(20 <
m_{\rm 814W} < 22)$ stars from  the Panchromatic Hubble Andromeda
Treasury (PHAT) survey, radial velocities of
over $5000$  red giant branch stars in the same magnitude range from
the Spectroscopic and Photometric Landscape of Andromeda's Stellar Halo
(SPLASH) survey, and integrated $I$-band SB profiles from
various sources. We use an affine-invariant Markov chain Monte Carlo algorithm to fit an
appropriate toy model to these three data sets. The bulge, disk, and halo
SB profiles are modeled as a S\'ersic, exponential, and cored power-law,
respectively, and the LFs are modeled as broken
power-laws.  We present probability distributions for each of 32 parameters
describing the SB profiles and LFs of the three
subcomponents. We find that the number of stars with a disk-like LF is
$5.2 \pm 2.1\%$ larger than the the number with disk-like (dynamically
cold) kinematics,
suggesting that some stars born in the disk have been dynamically
heated to the point that they are kinematically indistinguishable from
halo members. This is the first kinematical evidence for a ``kicked-up
disk'' halo population in M31. The fraction of kicked-up disk stars is
consistent with that found in simulations. We also find evidence for a
radially varying disk LF, consistent with a negative metallicity
gradient in the stellar disk. 

\end{abstract}
 
\section{INTRODUCTION}\label{intro_sec}

Spiral galaxies like the Milky Way (MW) and Andromeda (M31) are
composed of several structural subcomponents,
each with its own formation history. Tracing the
evolution of such a galaxy, then, depends on characterization of the
individual subcomponents. This decomposition is especially difficult in the inner
regions, where the cospatial bulge, disk, and halo complicate 
characterization of any single subcomponent. 

The most commonly used way to characterize the
stellar component of a large spiral galaxy is via surface brightness
(SB) decomposition. SB profiles of disks tend to
fall off exponentially, whereas bulges follow more general
S\'ersic profiles \citep{ser68}. Widely used SB decomposition codes
such as GIM2D \citep{sim02} and GALFIT \citep{pen02} fit a sum of a
S\'ersic spheroid and an exponential disk to the SB distribution of a
galaxy to estimate the relative contributions of the
components. Unfortunately, such fitting is plagued by degeneracies
that arise because the different subcomponents are cospatial and
because the procedure generally relies on ad hoc fitting formulae that
do not necessarily separate the galaxy into dynamically distinct
subcomponents \citep[e.g.,][]{aba03}.

In this paper, we present a new technique for decomposing the
stellar component of M31 into distinct bulge, disk, and halo
subcomponents. This technique can be applied to any galaxy close
enough to measure radial velocities and a luminosity function of stars
down to about 1.5 magnitudes fainter than the tip of the red giant branch
(TRGB). In addition to fitting
a toy model (exponential disk, powerlaw halo, and S\'ersic bulge) to
the SB distribution, we attempt to break the aforementioned
degeneracies by including an additional constraint: the fraction of
stars that belong to the disk (``disk fractions''), as measured from
stellar kinematics of individual red giant branch (RGB) stars. We
perform the decomposition using Bayesian techniques so that we can
identify covariances between model parameters, quantifying
lingering degeneracies. 

Two large-scale, ongoing resolved
stellar population surveys of M31 make it an ideal galaxy in which
to develop and test our decomposition technique. The
Spectroscopic and Panchromatic Landscape of Andromeda's Stellar Halo
(SPLASH) survey has used the Keck/DEIMOS multiobject spectrograph to
measure radial velocities of over 15,000 stars in M31
\citep{guh05,guh06}, including over 10,000 in the crowded inner 20 kpc
\citep{dor12,gil12,how13}. Meanwhile, the Panchromatic Hubble
Andromeda Treasury (PHAT) survey, a five-year Hubble
Space Telescope/Multi-Cycle Treasury (HST/MCT) program, has so far
imaged over $10^8$ stars in six filters in the UV, optical, and
near-IR in the same quadrant of the disk most densely sampled by
SPLASH \citep{dal12}. 

The decomposition technique presented here builds directly on
the work of \citet{cou11} and \citet{dor12}. The former performed
SB-only decompositions of M31 using two different techniques: Markov
chain Monte Carlo (MCMC) sampling and a nonlinear least-squares (NLLS) fitting
method.  In each case, the authors fit a sum of S\'ersic bulge, exponential disk, and
powerlaw halo to I-band SB profiles of the galaxy. The profiles were
a composite of major- and minor- axis cuts of an $I$-band image of the central
$\sim 20$ kpc of M31 \citep{cho02} plus minor-axis profiles measured
from RGB star counts \citep{pri94,irw05,gil09}. They found that the
best-fit bulge, disk, and halo structural parameters depended significantly on the
decomposition method, data points used (major vs. minor axis), and
even on the binning of the
data. Despite the uncertain results, their model was still relatively
restrictive,  not allowing for a flattened halo or
for variations from the canonical values in the ellipticities or
position angles of the bulge and disk. 

In \citet{dor12}, we performed a {\em kinematical} decomposition of
M31's dynamically cold disk (without distinguishing between thin and
thick components)
and dynamically hot spheroid (without distinguishing between bulge
or halo components) using
radial velocity measurements of $\sim 6000$ bright ($20 < I < 22$) RGB
stars in the inner parts of the galaxy ($5 < R_{\rm proj} < 20$ kpc;
the region dominated by the disk in optical and UV images). In each of $24$ small
spatial subregions, we measured the fraction of stars that belonged to
the hot component. This fraction is nonzero ($>10\%$) everywhere in
the survey region. The origin of this dynamically hot population is
unclear a priori: is it an outward extension of the central bulge or
the innermost reaches of the extended halo? On one hand, the bulge
appears too small in SB decompositions to contribute stars past a few
kpc \citep{cou11}; on the other hand, CMD analyses of fields between
$11$  and $45$ kpc on the minor axis reveal populations with young/intermediate
ages and intermediate metallicities \citep{bro03,ric08} ---
quite unlike the old, metal-poor halo field stars in the MW
\citep[e.g.,][]{car07,kal12}. 

In this paper, we use kinematically derived disk fractions
measured as in \citet{dor12} as constraints in a SB decomposition of
M31's $I$-band SB profile. Simultaneously fitting to the total SB profile of the
galaxy and the fraction contributed by the disk may help to reduce
degeneracies in the best fit parameters --- and possibly understand the
structural association and origin of the dynamically hot population
identified in \citet{dor12}. With more constraints, we can relax some
of the assumptions made in \citet{cou11}, fitting for the
ellipticities and position angle of the bulge, disk, and halo. 

A primary challenge in our work is that our disk fraction measurements
come from star count data, whereas we model the contributions to the
integrated surface brightness. Thus, we must convert a disk fraction
in star counts 
to a disk fraction in SB units. In general, this conversion factor is
such that the kinematical survey undersamples the disk: the
SPLASH target selection function happens to peak at magnitudes where
the spheroid and disk both contribute stars, and to fall off at
brighter magnitudes where only the disk contributes light. In
addition,  while it is easily recoverable, the
selection function is somewhat arbitrary, varying across the
galaxy. In order to quantify how fairly it
samples the three subcomponents, we model and fit for the disk,
bulge, and halo luminosity functions (LFs). 

The model presented here incorporates stellar population, kinematics,
and SB data for the first time. Including complexities like the boxy
bulge \citep{bea07} or separate thin,  thick, and extended disks
\citep{col11, iba05} are beyond the scope of the paper, which
nonetheless our analysis represents a significant improvement over
using only SB profile fits to characterize a galaxy's physical components. 

This paper is organized as follows. First, in \S\,\ref{sec_roadmap},
we outline the analysis procedures that will be described in detail in
\S\,\ref{sec_data}-\ref{sec_analysis}. In \S\,\ref{sec_data} we
summarize the kinematical, integrated-light, and resolved stellar
photometric data sets used as constraints. In \S\,\ref{sec_analysis}
we present the mathematical
formalism used to find the probability distribution functions (PDFs)
of each of the model parameters. In \S\,\ref{sec_results} we discuss
the probability distributions of our parameters, comparing them to
previous measurements of M31's structural parameters. In
\S\,\ref{sec_discussion}, we show that the dynamically hot population
in our kinematical sample is more closely associated with the extended
halo than the central bulge, and propose that some fraction of that
population originated in the disk. Finally, we summarize in
\S\,\ref{sec_summary}.

\section{Overview of Analysis Procedure}\label{sec_roadmap}

Simultaneously fitting even a simple toy model to the observed SB map,
disk fractions, and LF is relatively complicated. We must process
three different data sets, develop a multiparameter model, combine the
model and data into a likelihood function, and sample that
likelihood to obtain a probability distribution function
(PDF) for each parameter. Each of these steps will be described in
detail in \S\,\ref{sec_data}-\ref{sec_analysis}, but first we will
present a brief road map to those sections. 

\begin{figure*}
\scalebox{0.6}{\includegraphics[trim=70 270 170 70, clip =
  true]{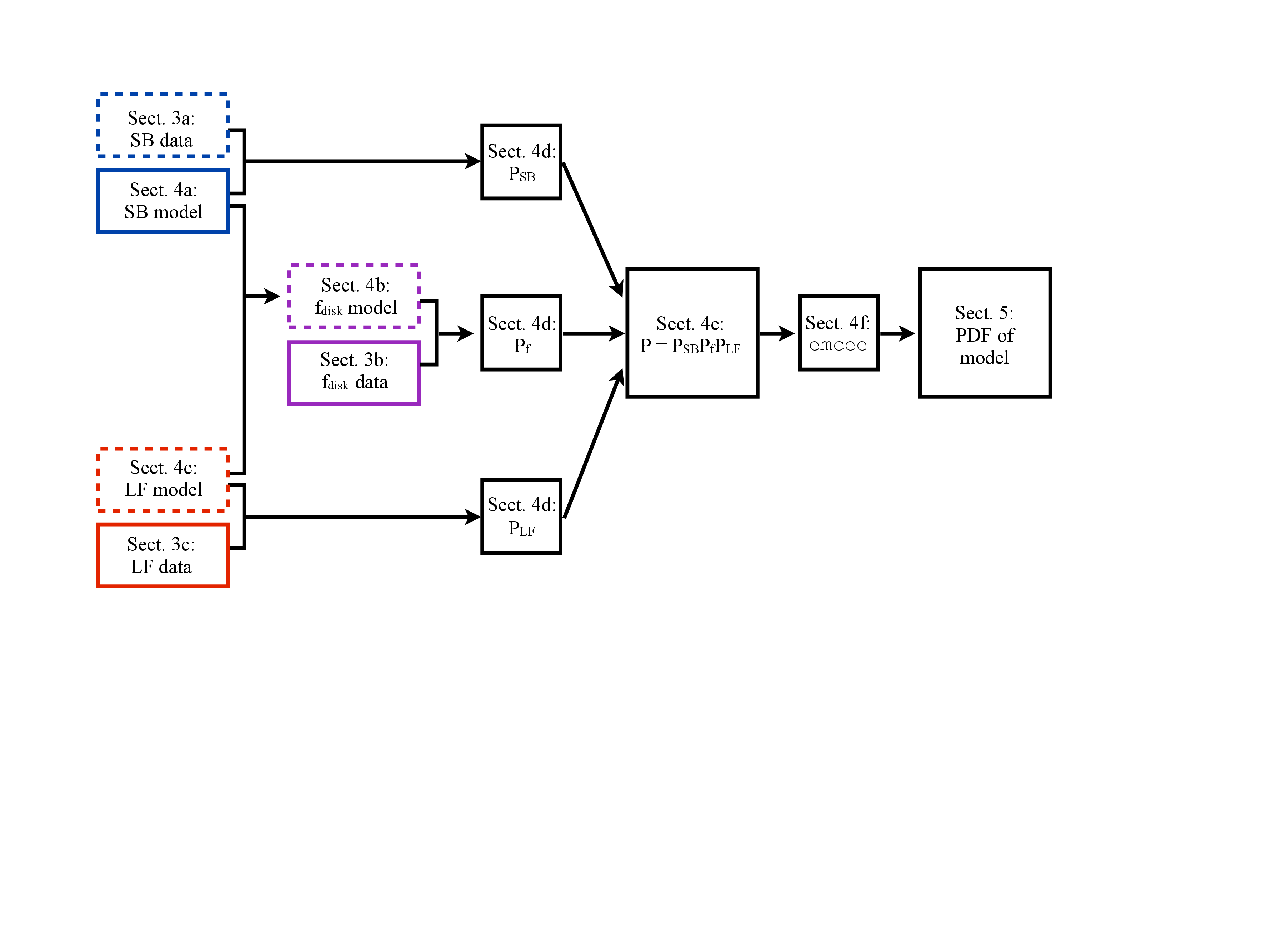}}
\centering
\caption{Flow chart illustrating the analysis procedure. Each box is labeled with
  the section of the text in which that step is described. See text of
  \S\,\ref{sec_roadmap} for description.}
\label{fig_flowchart}
\end{figure*}

The flow chart in Figure~\ref{fig_flowchart} illustrates the path from
data processing to model parameter PDFs. We start with three sets of
observational constraints: SB profiles, the kinematically-derived disk
fraction probability distribution in each of the spatial subregions,
and the $F814W$ ($I$) LF in each subregion. The
processing of these three data sets is described in
\S\,\ref{ssec_phot}-\ref{ssec_lf}. 

As the left-hand side of Figure~\ref{fig_flowchart} shows, each data
set has an accompanying toy model. We model the SB as the sum of
three profiles: a S\'ersic bulge, an exponential disk, and a cored
power-law halo (\S\,\ref{sssection_profiles}). Similarly, we model the
luminosity function as a sum of three broken powerlaws, one each for
the disk, bulge, and halo, in the magnitude range $m_{\rm F814W}=[20,
22]$ (\S\,\ref{sss_modellf}). The
model for the disk fraction comes from a combination of the previous
two: the disk fraction in flux units at a given location is 
the ratio of the disk to total SB at that location; however, we must
convert this integrated-light disk fraction to star counts using the
disk LF (\S\,\ref{sssection_modeldiskf}). 

Given a certain vector $w$ in parameter space, we compute the
probability $P_{\rm SB}(w)$ that the SB model represents the observed
SB map; similarly, we define and compute $P_{\rm f}(w)$ and $P_{\rm
  LF}(w)$ as the probabilities that the model matches the kinematical
and LF data sets, respectively. The total probability of that vector
$w$ is then $P(w) = P_{\rm SB}P_{\rm   f}P_{\rm
  LF}$. \S\,\ref{ssec_likelihood} gives the equation for $P$. 

We then sample this probability distribution $P$ using the Markov
chain Monte Carlo (MCMC) sampler {\tt emcee} \citep{for12}, as
described in \S\,\ref{ssec_mcmc}. The sampler yields a PDF of each
model parameter. In \S\,\ref{sec_results}, we discuss the median
values and confidence intervals of the PDFs in the context of previous
measurements. 


\section{Observational Constraints}\label{sec_data}
The integrated-light, resolved stellar photometric, and spectroscopic
data sets used in this work are described in more detail in
\citet{cou11}, \citet{gil12}, \citet{dal12}, and \citet{dor12}. In
this section we briefly recap the salient information. 

\subsection{Surface Brightness}\label{ssec_phot}
The SB map is derived from two classes of data: major- and minor-axis wedge cuts
from a mosaic image, and RGB star counts. All of the profiles are in
the $I$ band to minimize the effects of inhomogeneities such as dust
lanes or the UV-bright star-forming rings.  The SB data used in this
work are identical to
those used in \citet{cou11}, with the addition of new halo fields out
to a galactocentric projected radius of $R=225$ kpc first presented in \citet{gil12}. 

\subsubsection{Integrated Light Profiles}
We use the major- and minor-axis SB profiles of M31 that \citet{cou11}
constructed from an $I$-band CCD mosaic image of M31 taken with the Kitt
Peak National Observatory Burrell Schmidt telescope and presented in
\citet{cho02}. Rather than relying on azimuthally
averaged isophotal fitting, which smears together components with very
different ellipticities such as bulge and disk, they cut wedges from
the major and minor axes of the galaxy. They
measured the median SB in each of 186 bins on the major axis and 152
on the minor axis.  The sizes of the bins increased with
galactocentric radius. Bright stars were removed via an iterative
sigma-clipping process. The quoted error bar in each bin is the RMS
deviation about the median value of the pixels in that bin. The
locations of these bins on the sky are marked as teal circles in
Figure~\ref{fig_phot} and in Figure~\ref{fig_zoomphot}, the latter
zooming in on the the portion of the galaxy dominated by the bright
disk and bulge. 

Because the CCD pixel size is $2''$ and the telescope was not well focused during
some of the observations, some of the images that went into the
final mosaic image are out of focus at the $2''-5''$ level. (See \S\,2
of \citet{cho02}.) To avoid
the effects of finite spatial resolution, we exclude from our analysis
the two bins within projected radius $R = 10''~(40 ~\rm pc)$ of the
center of M31.

\subsubsection{Star Counts}

In the outer regions of the galaxy, where crowding is less severe, the
SB profile
can be measured, up to a normalization factor, from RGB star counts. 
Like \citet{cou11}, we have combined the extended
star counts of the M31 stellar halo by
\citet{pri94}, \citet{irw05}, \citet{gil09}, and \citet{gil12}. These
profiles cover the range $20~{\rm kpc} < R < 225$ kpc.

\begin{figure}[h]
\scalebox{0.6}{\includegraphics[trim=65 5 90 35, clip =
  true]{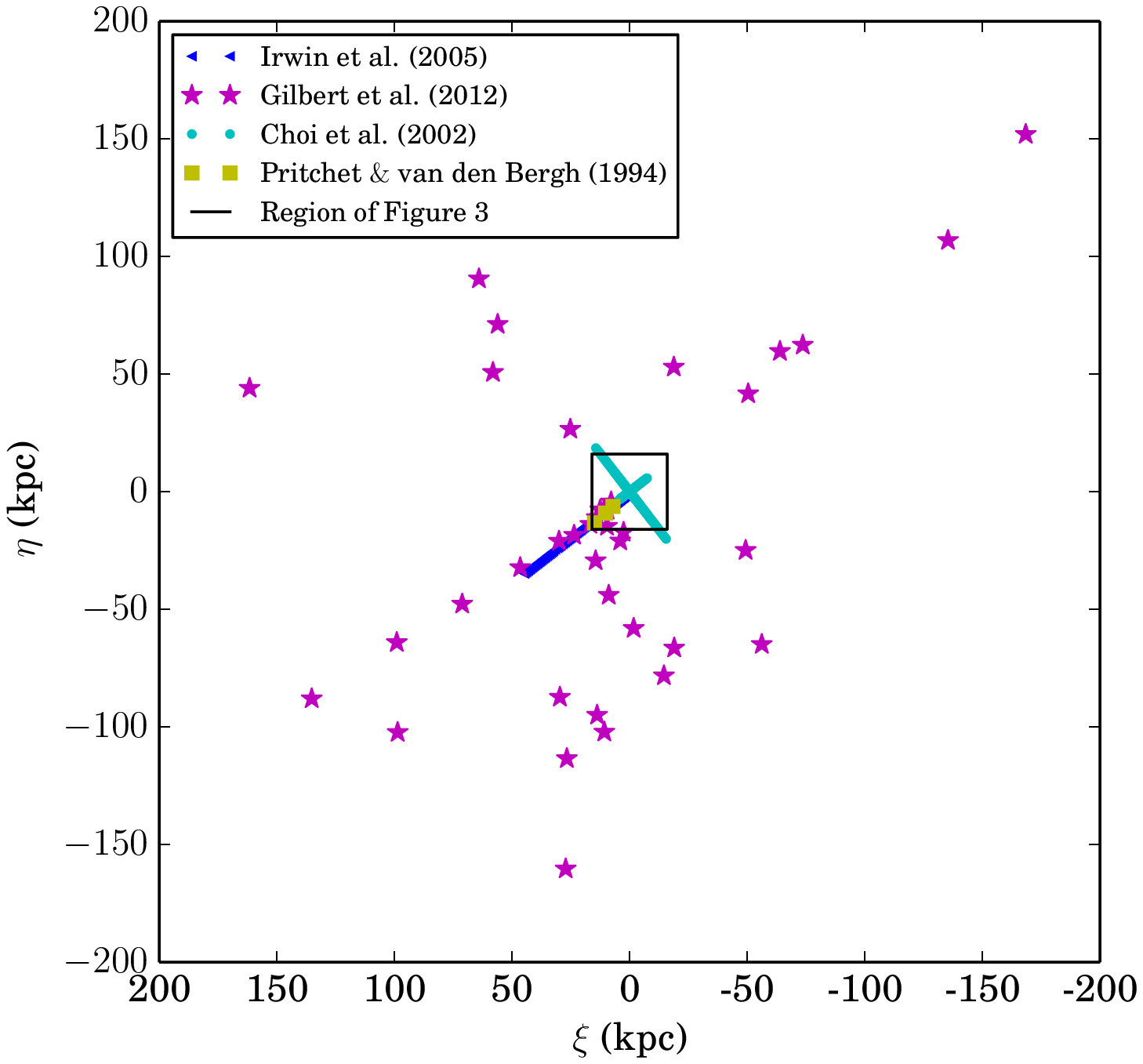}}
\centering
\caption{Locations of 637 fields from 4 datasets used for SB measurements. The
  \citet{cho02} measurements (teal circles) come from a wedge cut of an I-band
  image from KPNO; the rest of the SB measurements, including
  \citet{irw05} (blue triangles), \citet{gil12} (violet stars), and
  \citet{pri94} (yellow squares) are derived from
  RGB star counts. The inner black
  box has a side length of 32 kpc, approximately the size of the GALEX
image of M31. A zoom-in of this region is shown in Figure~\ref{fig_zoomphot}.}
\label{fig_phot}
\end{figure}

\begin{figure*}[h]
\scalebox{1}{\includegraphics[trim=65 5 100 45, clip =
  true]{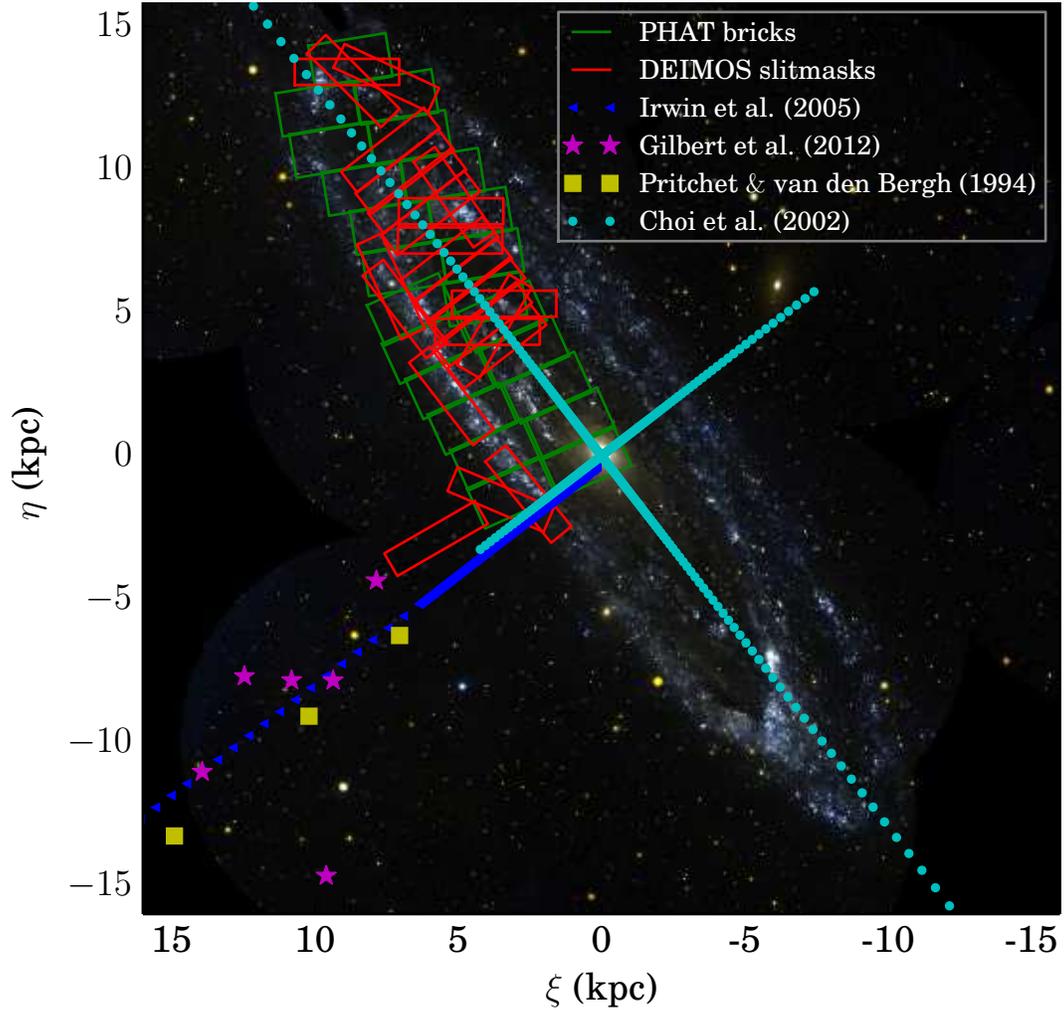}}
\centering
\caption{Zoom-in on the map of SB fields from
  Figure~\ref{fig_phot} that fall in the inner regions of
  M31, overlaid on a GALEX image. Points are color-coded as in
  Figure~\ref{fig_phot}. The \citet{irw05} points (blue triangles) are drawn slightly
  offset from the minor axis for
  clarity; they actually overlap the \citet{cho02} minor axis
  points (teal circles). For comparison, bricks in the HST PHAT survey (from which we
  measure the bright end of the luminosity function) are shown in
  green and Keck/DEIMOS slitmasks from the SPLASH survey (from which
  we measure radial velocities of RGB stars) are outlined in red. The
  kinematical and LF analyses presented in this paper are carried out
  in the regions where the green bricks and red slitmasks overlap.}
\label{fig_zoomphot}
\end{figure*}

The \citet{pri94} data set --- which combines digital star counts to
measure the SB in three fields along the minor axis of the galaxy ---
is identical to that used in \citet{cou11}. These fields are marked in
yellow squares in Figures~\ref{fig_phot} and \ref{fig_zoomphot}.  

The \citet{irw05} data set is also identical to that used in
\citet{cou11}. \citet{irw05} combined the data from \citet{pri94} with faint RGB star
counts to trace the minor-axis stellar distribution out to a projected
radius of $55$ kpc. They used star counts from the Isaac Newton
Telescope, exposing typically $800-1000$ s per field
in the Gunn $i$ band. These fields are shown as blue triangles in
Figures~\ref{fig_phot} and \ref{fig_zoomphot}.

The final SB data set, presented in \citet{gil09} and \citet{gil12}, is based on RGB
star counts from a Keck/DEIMOS spectroscopic survey of 38 fields
between $9-225$ kpc in projected radius [rather than just the 12
fields used by \citep{cou11}]. Members of M31's smooth halo
were identified and distinguished from foreground MW dwarf
star contaminants and substructure in the M31 halo
using a combination of photometric and spectroscopic diagnostics
\citep{guh06, gil06}. To estimate M31's stellar surface density as a
function of radius, the observed ratio of M31 red giants to MW dwarf stars was multiplied by the surface density of MW
dwarf stars predicted by the Besan\c{c}on  Galactic star-count model
\citep[e.g.,][]{rob03,rob04}. The surface density was converted to
SB units by scaling to match the minor axis profile from \citet{cho02}. These fields are
shown as magenta stars in Figures~\ref{fig_phot} and
\ref{fig_zoomphot}. 

In summary, we have 637 SB measurements $\mu_{\rm obs}$ from 4 data
sets. Though the SB data are given in magnitudes, we perform the SB fits
using flux units; the conversion is given through the usual formula: 

\begin{eqnarray}\displaystyle
  \Sigma_{\rm obs}(R, \Delta PA) = 10^{-(\mu_{\rm obs}+ zp)/2.5}
  \label{eq:lum_counts}
\end{eqnarray}

\noindent where $\Sigma_{\rm obs}$ is in flux
units, $\mu_{\rm obs}$
is in magnitudes, and $zp$ is the zeropoint. To ensure consistency between
datasets, we anchor the zeropoints of the \citet{pri94, gil12}, and \citet{irw05}
datasets to that of the \citet{cho02} image by first fitting to
the  composite SB map {\em only} (i.e., not including constraints on the
disk fraction), leaving the zeropoint as a free parameter with a flat
prior between $\pm 2$ magnitude of the \citet{cho02} value. For each
dataset, the median (best fit) zeropoint of the resulting posterior
distribution is adopted for the rest of the
analysis. 

\begin{figure*}
\scalebox{0.7}{\includegraphics[trim=00 0 0 30, clip =
  true]{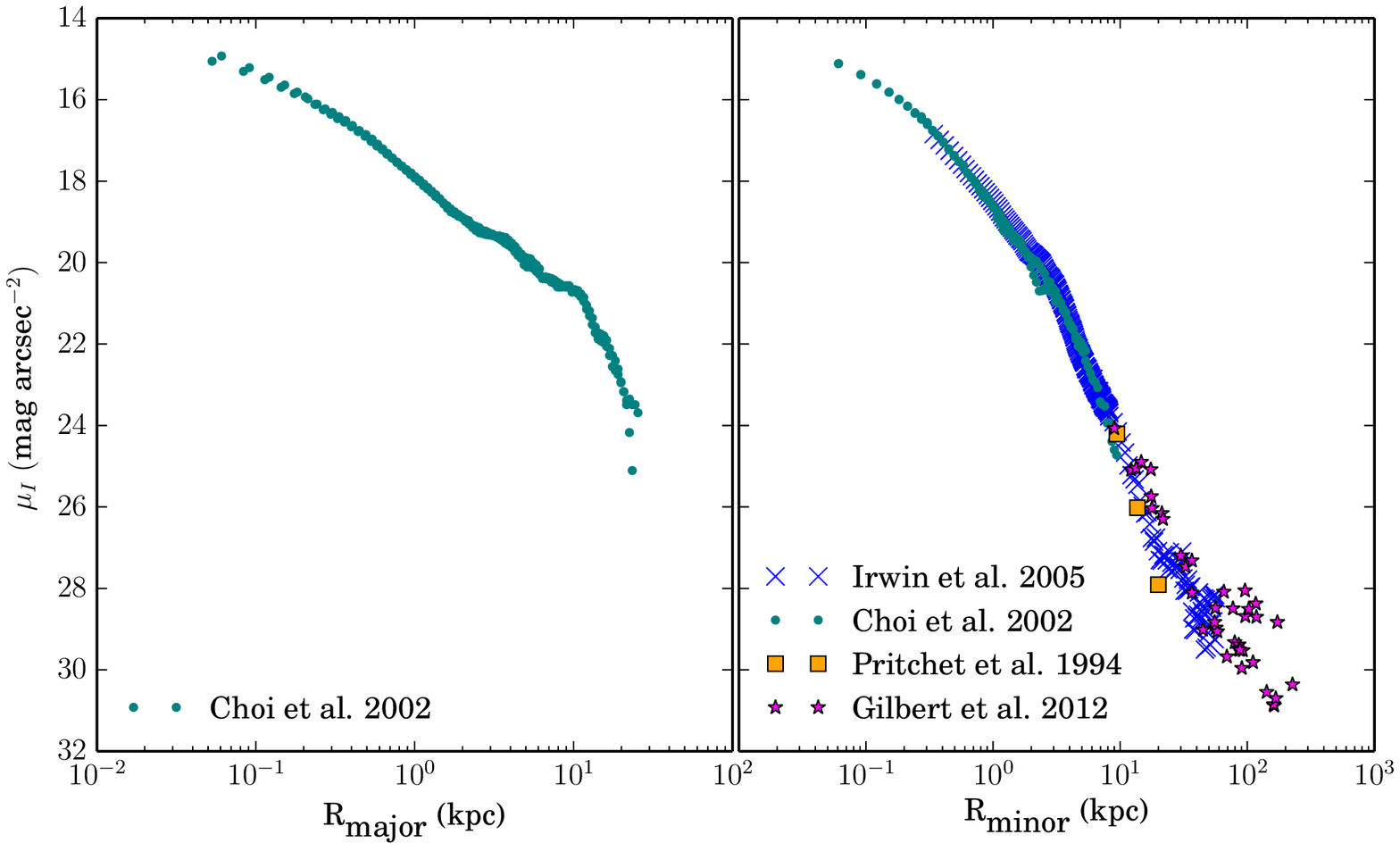}}
\centering
\caption{Major- (left) and minor- (right) axis projections of the SB
  profiles, assuming a major axis position angle of
  $37.7^{\circ}$. Samples from opposite sides of the galactic center
  are collapsed onto a single set of axes. \citet{gil12} points (violet
  stars) do not lie on
  either axis,  so for display purposes we project them to the minor
  axis assuming a  circularly symmetric halo. Photometric
  uncertainties are typically   smaller than the point size.}
\label{fig_obssb}
\end{figure*}

The composite, zeropoint-adjusted major- and minor-axis SB profiles
are shown in the two panels of Figure~\ref{fig_obssb}.

\subsection{Velocity distribution}\label{ssec_fdisk}
The constraints on the fraction of stars dynamically associated with
the disk are obtained via a kinematical decomposition of the stellar
velocity distribution into dynamically cold (disk) and hot (spheroid,
or combined bulge and halo) components, described in detail in
\citet{dor12}. 

\subsubsection{Kinematical data}
We obtain spectra of 5257 stars in
the inner 20 kpc of M31 using the DEIMOS multi-object spectrograph on
the Keck II telescope as part of the SPLASH survey
\citep{guh05,guh06,gil09,dor12,how13}. The positions
of the 29 DEIMOS slitmasks are shown as red rectangles overlaid on a
GALEX image of M31 in Figure~\ref{fig_zoomphot}. The targets on 19 of
these masks contain targets selected from an $i'$-band CFHT/MegaCam
image, while the targets on 5 masks were selected exclusively from the
PHAT catalog and the targets on the remaining 5 were selected using a
combination of PHAT and CFHT data. In each case, the
targets are selected based on a complex set of factors: apparent
degree of isolation in the source catalog, magnitude bright enough to
yield a high-S/N spectrum but not so
bright as to be a likely foreground MW dwarf, and position on the sky
so as to optimize the use of the slitmask. Hence, the luminosity
function of DEIMOS-observed stars is different in each slitmask, but
in general is nonzero between $20 < i^{\prime} < 22$ and peaks
somewhere around $i^{\prime} \sim 21$. 

Each spectrum is collapsed in the spatial direction and
cross-correlated against a suite of template rest-frame spectra to
measure its radial velocity. The velocity measurement is then
quality-checked by eye, and only robust velocities (those based on at
least two strong spectral features) are included in the analysis. 

We perform a number  of cuts on this preliminary kinematical dataset. First, we
retain only the 3247 (63\% of) targets that are also present in the
existing PHAT catalog. (The other targets lie in regions on the sky
not sampled by the PHAT survey at the time of writing.) Second, we cut
out the 25\% of the remaining objects that are located in regions with
$A_v > 1.0$ mag. The low extinction regions were identified by
modeling the NIR CMD as a sum of a foreground unreddened RGB and a
background RGB, reddened by a
log-normal distribution of dust reddening, in 10 arcsecond bins.  This
modeling produced a map of the median $A_V$ across the disk,
independent of the fraction of reddened stars along a given line of
sight. Details can be found in Dalcanton et~al. (2013, in prep), and
an overview of the technique can be found in \citet{dal12}. All
of the 2443 remaining stars are redder than $m_{\rm F475W} - m_{\rm
  F814W} = 1.5$, so are likely to be RGB stars rather than bright main
sequence (MS) objects. In the following text, we describe how we use
these 2443 stars to measure disk fractions in each of 14 spatial
subregions. 

\subsubsection{Review of Disk Fraction Measurements}\label{sec_diskf}

In \citet{dor12}, the SPLASH and PHAT survey region was divided into
four spatial ``regions'': three straddling the NE major axis and one along
the SE minor axis. The boundaries of the regions (lines of constant
$R_{\rm proj}$ or position angle P.A.) were chosen such that each region contained
enough SPLASH targets to constrain the velocity distribution of the
subdominant spheroid component. Each region was then subdivided along
lines of constant P.A. into multiple ``subregions,'' each large enough
to contain $>100$ stars but small enough that the rotation velocity of
the disk component does not vary substantially.  We measured the mean
velocity $v$, velocity dispersion $\sigma_v$, and fraction $f$ of the disk and
spheroid components in each of the resulting 24 subregions by fitting a sum
of two Gaussians (representing a dynamically cold disk
and dynamically warmer spheroid)  to the velocity distribution in each
subregion. 

\begin{figure*}
\scalebox{1}{\includegraphics[trim=65 300 50 300, clip =
  true]{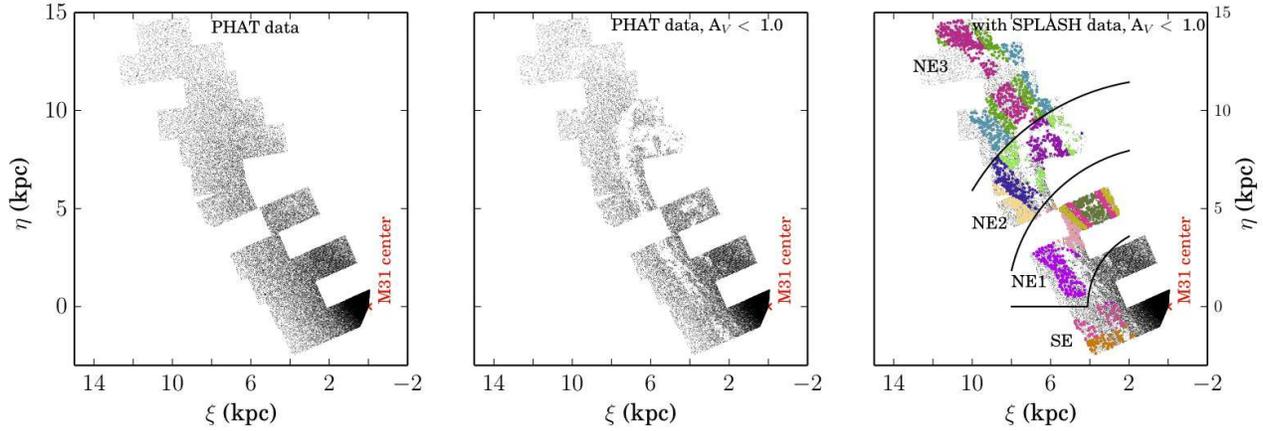}}
\centering
\caption{Map of target locations in the PHAT and SPLASH surveys. {\em Left:}
Stars in the magnitude range $m_{F814W}=[20,22]$ in PHAT. (Only a random
4\% of the objects are shown.) {\em
  Middle:} Same as left panel, but without stars in dusty ($A_v >
1.0$) pixels. The star-forming $10$ kpc ring is effectively
excluded. {\em Right:} Same as middle panel, with stars from the SPLASH
survey in non-extincted regions overplotted as colored dots. Black
lines separate the 4 large ``regions,'' while colors demarcate
the 14 smaller ``subregions.'' The $F814W$ luminosity functions
and kinematically-derived disk fractions are measured in each
subregion, where the kinematical and PHAT surveys overlap. Note that
the central few kpc are not used in the fits to the kinematics and LF,
because this region is too crowded for resolved stellar spectroscopy
with DEIMOS.}
\label{fig_datamap}
\end{figure*}

Because in the current paper we use only the subset of
spectroscopic targets also
identified in the PHAT survey, we must combine some subregions from
\citet{dor12} to have sufficient number statistics 
to reliably separate the disk and spheroid. The 14 subregions we use are
shown in the right-hand panel of Figure~\ref{fig_datamap}. The
subregions are named using the same formalism as in \citet{dor12}. The
survey area is still divided into four regions: SE (along the SE minor axis), and NE1,
NE2, and NE3 (along the NE major axis, in order of increasing projected
radius). The subregions within each region are
identified with subscripts that increase with distance from the major
axis: $\rm NE1_1, NE1_2, \ldots, NE1_5;~NE2_1, \ldots, NE2_4$, and
  $\rm NE3_1,NE3_2,NE3_3.$ In the SE
region, the inner south, and inner north subregions are named
$\rm SE_2$ and  $\rm SE_3,$ respectively. (Subregion $SE_1$ is not
used because it does not overlap with the PHAT survey region.)

We fit a sum of two Gaussians (representing a dynamically cold disk
and dynamically warmer spheroid) to the velocity distribution in each
subregion with the MCMC sampler {\tt emcee} \citep[see the Appendix
for more information]{for12}. We allow the disk velocity distribution
to change from subregion to subregion, but require that
the mean velocity and dispersion of the spheroid remain constant between
subregions within a region. The decomposition makes almost no
assumptions about the nature of the disk, except that its velocity
distribution is symmetric and locally colder than that of the
spheroid.

 The latter is perfectly reasonable; however, one could
imagine that asymmetric
drift in a stellar disk could skew the distribution of circular
velocities towards lower speeds and upweighting the spheroid in the
decomposition. We found that asymmetric drift does {\em not} in
fact have a significant effect on the kinematically derived disk
fractions. We tested the effects of asymmetric drift using the
velocity distribution presented
in \citet{sch12} to measure disk fractions in the major axis
subregions, where the effect of asymmetric drift should be most significant. In
near-major-axis subregions ($NE1_{1-3}$, $NE2_{1-2}$,
$NE3_{1-3}$), the disk fractions computed by assuming symmetric and
asymmetric velocity distributions are consistent to within $1\sigma$. In subregions
that are closer to the minor axis, it is difficult to fit the
asymmetric function to the velocity distribution because the 
line-of-sight velocity measurements do not well constrain the rotation
velocity. However, for the same reason, we expect the effects of
asymmetric drift on the measured disk fraction to be decrease towards
the minor axis. The line-of-sight velocity of a star on
the minor axis is independent of the circular velocity of that star,
and so a slight change in the circular velocity distribution should
not affect the line-of-sight distribution. Therefore, we
can safely ignore the effect of asymmetric drift and use a Gaussian
disk velocity distribution without affecting the disk fraction measurements.

\begin{figure*}
\scalebox{1.0}{\includegraphics[trim=30 140 0 0, clip =
  true]{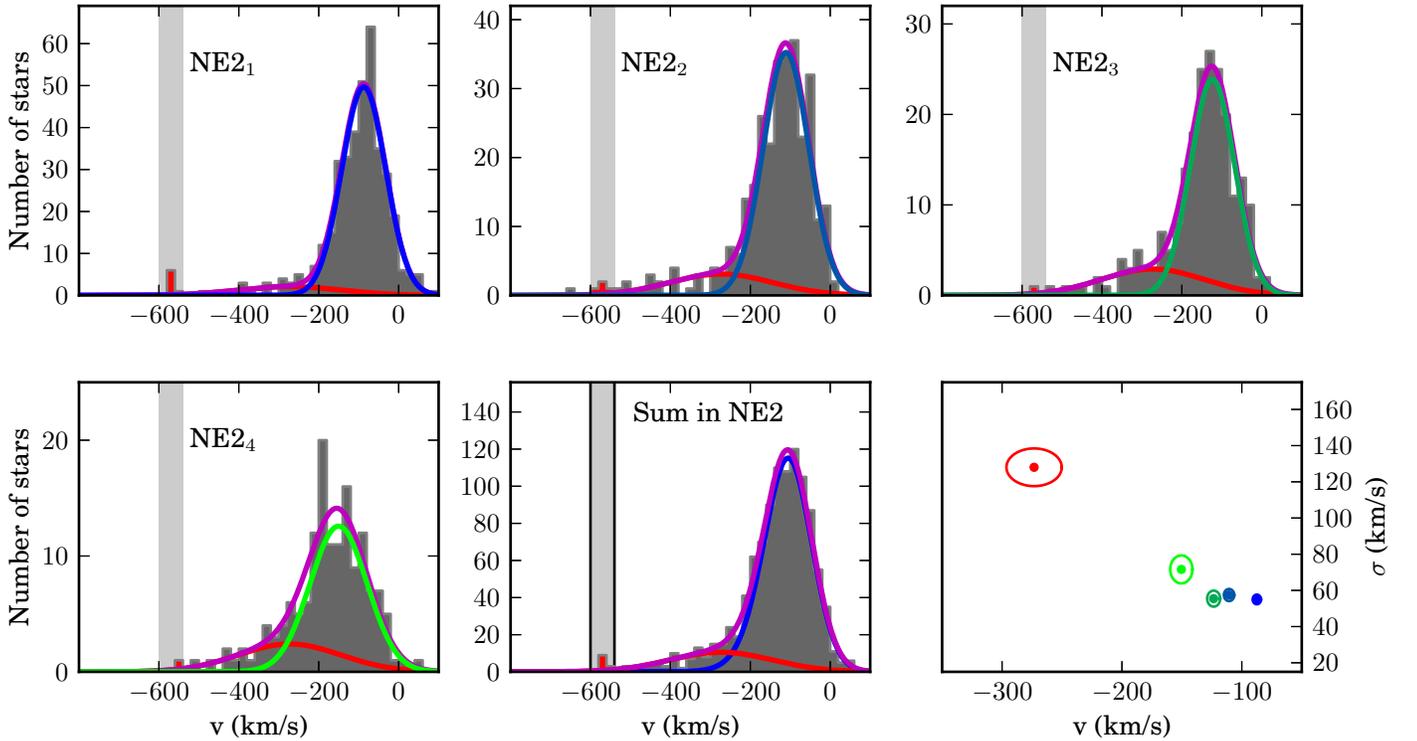}}
\centering
\caption{{\em First four panels:} fits to the velocity distribution of RGB
  stars from the SPLASH survey in each of four subregions in the NE2
  (middle northeast major axis) region, plotted over velocity
  histograms of stars in each subregion. Each velocity distribution is
  fit by a sum of two Gaussians corresponding to a dynamically hot
  spheroid (red) and dynamically cold disk (blue to green). The sum of the two
  curves is shown in violet. Velocity ranges excluded from the fits
  due to possible contamination by tidal debris from the Giant
  Southern Stream (GSS) are shown
  in two equivalent ways: by the gray shaded regions and by the stars
  shaded red on the histogram. {\em Bottom middle panel:} Sum of hot
  components in the NE2 region (red curve) and sum of cold components
  (blue curve)  overplotted on a histogram of all radial velocities in
  NE2. {\em Bottom right panel:} Ellipses represent the mean and
  uncertainty of each of the parameter pairs $(v, \sigma_v)$. The four
  blue and green ellipses represent the kinematical parameters of the
  disk components in the four subregions. The red ellipse respresents
  the spheroid component, which has the same $(v, \sigma_v)$ in
  every subregion within this region.}
\vspace{5mm}
\label{fig_dor12}
\end{figure*}

Figure~\ref{fig_dor12} illustrates the velocity decomposition in each
subregion within a representative region, NE2, which is centered along the
northeast major axis at a projected radius of about 10 kpc. Note that in each
subregion within the NE2 region, some nonzero fraction of stars belong
to the dynamically hot component. Nearly half of these spheroid stars are {\em
  counterrotating} relative to the disk (that is, they have radial
velocities less than $v_{\rm M31}=-300$ km/s), and thus must be
spheroid stars regardless of the detailed shape of the disk velocity
distribution. In \S\,\ref{sec_discussion}, we will show that
these stars are more closely associated with the extended halo than
the central bulge, and that some of them likely originated in the
disk. 

The cold fraction in each subregion $s$ is the
ratio of the areas of the cold Gaussian and the total velocity
distribution. At the end of the 10,000-step
MCMC chain, the disk fractions measured from the 32 walkers over the final
2000 steps of the chain are compiled and binned by $0.5\%$ into
a normalized PDF $p_s(f_{\rm   d})$. 

Because it will be more useful to have an analytic form for
$p_s(f_{\rm d})$, we fit a skew-normal function $\phi_s$ to
$p_s(f_{\rm d})$ in each subregion $s$ using
a Levenberg-Marquardt minimization. The form of $\phi_s$ is 

\begin{eqnarray}\displaystyle
 \phi_s(f) = A_se^{-(z_s/\sqrt{2})^2}\bigg(1 +
  {\rm erf}\bigg(\frac{\tau_s z_s }{\sqrt{2}}\bigg)\bigg)
  \label{eq:skew}
\end{eqnarray}

\noindent where

\begin{eqnarray}\displaystyle
  z_s = (f_s - \mu_s)/\sigma_s.
\label{eq:z}
\end{eqnarray}

\begin{figure}
\scalebox{0.5}{\includegraphics[trim=40 10 50 15, clip =
  false]{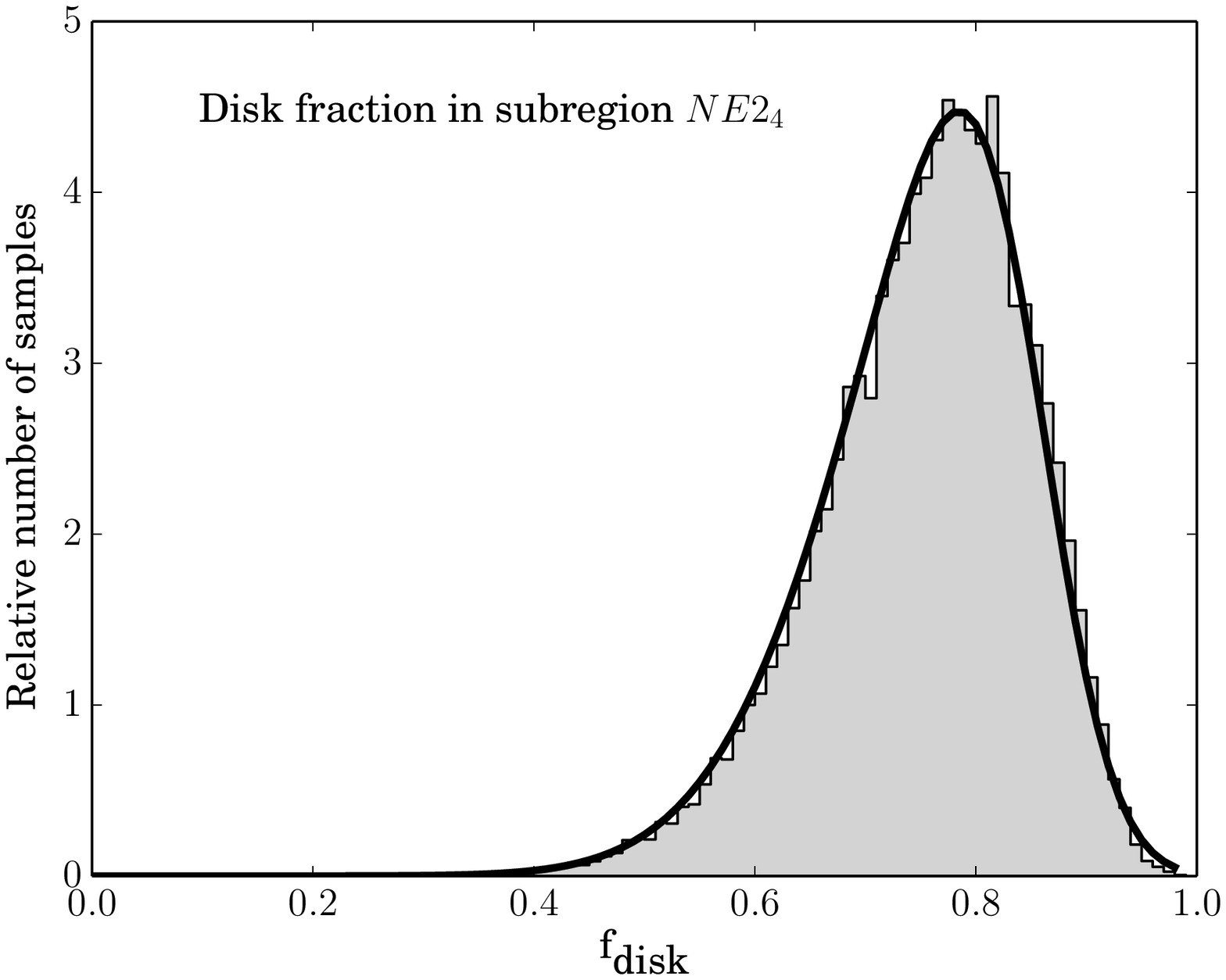}}
\centering
\caption{Probability distribution of the kinematically derived disk
  fraction in the off-major-axis subregion $\rm{NE2}_4$.  The gray
  filled histogram is constructed from 50,000 samples from the end of
  the MCMC chain used to fit the sum of two Gaussians (disk and
  spheroid) to the observed velocity distribution in this
  subregion. The black curve represents the skew-normal function $\phi$
  that best fits the histogram.  $\phi$ is later used as the prior on the
  disk fraction; that is, a model that predicts a disk fraction near
  the peak (e.g., $f_{\rm d} \sim 0.8$) will be more favorable than one
that predicts a disk fraction far from the peak (e.g., $f_{\rm
  disk} \sim 0.3$). }
\label{fig_skew}
\end{figure}

The best-fit skew-normal parameters $\mu_s, \sigma_s, \tau_s$, along
with the
medians and standard deviations of the $p_s$ distributions, are given in Table~\ref{tab_skew}.  Figure~\ref{fig_skew} shows an example fit to
$p_s(f_{\rm d})$ in representative subregion $\rm NE2_4$.  The
PDFs are very well approximated by a skew-normal function: the
$\chi^2~ p$-statistic is 1.000 in every subregion. The uncertainties in
the disk fraction measurements in each subregion are parameterized by
the width of the skew-normal
function; the entire function (not just an error bar) will be used as
a prior on the disk fraction in the decomposition presented. Note that
because the subregions are our smallest resolution elements in our
kinematical analysis, we cannot map the variation in
kinematically-derived disk fraction {\em within} a subregion. However, in
subregions where the true disk fraction varies significantly, the
PDF is broad --- in other words, systematic uncertainties in
the measured disk fractions derived from finite spatial binning are
incorporated into the error bars.  

\begin{deluxetable}{lllcccccccc}
\tabletypesize{\scriptsize}
\tablecaption{Skew-Normal Parameters of \\
Disk Fraction Probability Distribution $p_s(f_{\rm d})$}

\tablewidth{0 pt}
\tablehead{
\colhead{Subregion}&
\colhead{$\mu$}&
\colhead{$\sigma$}&
\colhead{$\tau$}&
\colhead{A}&
\colhead{Median}&
\colhead{Std. dev.}&
}
\startdata
${\rm{NE1_1}}$ & ~0.75 & ~0.11 & -1.64 & ~1.83 & 0.68 & 0.08 \\ 
${\rm{NE1_2}}$ & ~0.82 & ~0.11 & -1.73 & ~1.86 & 0.75 & 0.08 \\ 
${\rm{NE1_3}}$ & ~0.56 & ~0.11 & -0.78 & ~1.88 & 0.51 & 0.09 \\ 
${\rm{NE1_4}}$ & ~0.83 & ~0.22 & -1.65 & ~0.95 & 0.70 & 0.15 \\ 
${\rm{NE1_5}}$ & ~0.77 & ~0.10 & -1.11 & ~1.97 & 0.72 & 0.08 \\ 
${\rm{NE2_1}}$ & ~0.96 & ~0.04 & -2.43 & ~4.61 & 0.93 & 0.03 \\ 
${\rm{NE2_2}}$ & ~0.86 & ~0.08 & -2.04 & ~2.4 & 0.81 & 0.06 \\ 
${\rm{NE2_3}}$ & ~0.87 & ~0.08 & -2.00 & ~2.57 & 0.82 & 0.05 \\ 
${\rm{NE2_4}}$ & ~0.84 & ~0.15 & -2.53 & ~1.38 & 0.75 & 0.10 \\ 
${\rm{NE3_1}}$ & ~0.83 & -0.04 & ~1.47 & ~5.12 & 0.81 & 0.03 \\ 
${\rm{NE3_2}}$ & ~0.86 & ~0.05 & -1.35 & ~4.04 & 0.84 & 0.04 \\ 
${\rm{NE3_3}}$ & ~0.84 & ~0.06 & -1.34 & ~3.50 & 0.81 & 0.04 \\ 
${\rm{SE_2}}$ & ~0.30 & ~0.25 & -0.07 & ~4.53 & 0.33 & 0.21 \\ 
 ${\rm{SE_3}}$ & -2.68 & ~3.16 & -1.26 & 19.30 & 0.37 & 0.29 \\ 

\enddata
\label{tab_skew}

\end{deluxetable}

\subsection{Luminosity Function}\label{ssec_lf}
The bright end of M31's LF is the crucial link
between the kinematical and SB data. The kinematics measure the
fraction of stars, as sampled by the SPLASH survey, that belong
to the disk in each small subregion, while a SB decomposition yields the
fraction of integrated light contributed by the disk. To
convert between these units, we must know how fairly the SPLASH survey
samples the bulge, disk, and halo LFs. We therefore fit three model LFs
to the observed PHAT LF in each subregion. This section describes the
observed LF; the model LFs are discussed in \S\,\ref{sss_modellf}. 


We measure the PHAT LF in the magnitude range sampled by SPLASH ($20 <
m_{\rm F814W} < 22$). While the PHAT survey is  crowding-limited and thus
incomplete at faint magnitudes, within our radial range,$6 < R < 20$
kpc in deprojected radius, it is $100\%$ complete down to
$m_{\rm F814W} = 22$ for colors $m_{\rm F475W}-m_{\rm F814W} <
4.5$. (We have tried a decomposition excluding the few percent of stars
redder than this cutoff, but the qualitative results are not affected.) We clean
the sample in
the same way as for the SPLASH data set, using
 only stars redder than $m_{F475W}-m_{F814W}=1.5$ and in pixels with
 $A_v  < 1.0$ mag. We bin these stars by $0.1$ mag in the range $20 <
 m_{F814W} < 22$. Figure~\ref{fig_datamap} shows the PHAT stars in
 this magnitude range before (left) and after (right) excluding stars
 in extincted regions. 


In each subregion $s$, we measure two normalized luminosity
functions in the magnitude range sampled by the SPLASH spectroscopic
survey $(I = [20, 22])$. $\mathcal{L}_{\rm PHAT,s}$ contains stars from
the PHAT catalog that fall into subregion $s$. $\mathcal{L}_{\rm SPLASH,s}$
contains stars found in both the SPLASH and PHAT surveys in subregion
$s$. Each has units of number/$\textrm{arcsec}^2$/mag.  The
uncertainty on each observed LF is simply the Poisson counting
uncertainty on each bin $\sqrt{N}$, where $N$ is the number of stars
in the subregion that fall into that magnitude bin. 
 
\begin{figure*}
\scalebox{0.9}{\includegraphics[trim=0 150 20 40, clip =
  true]{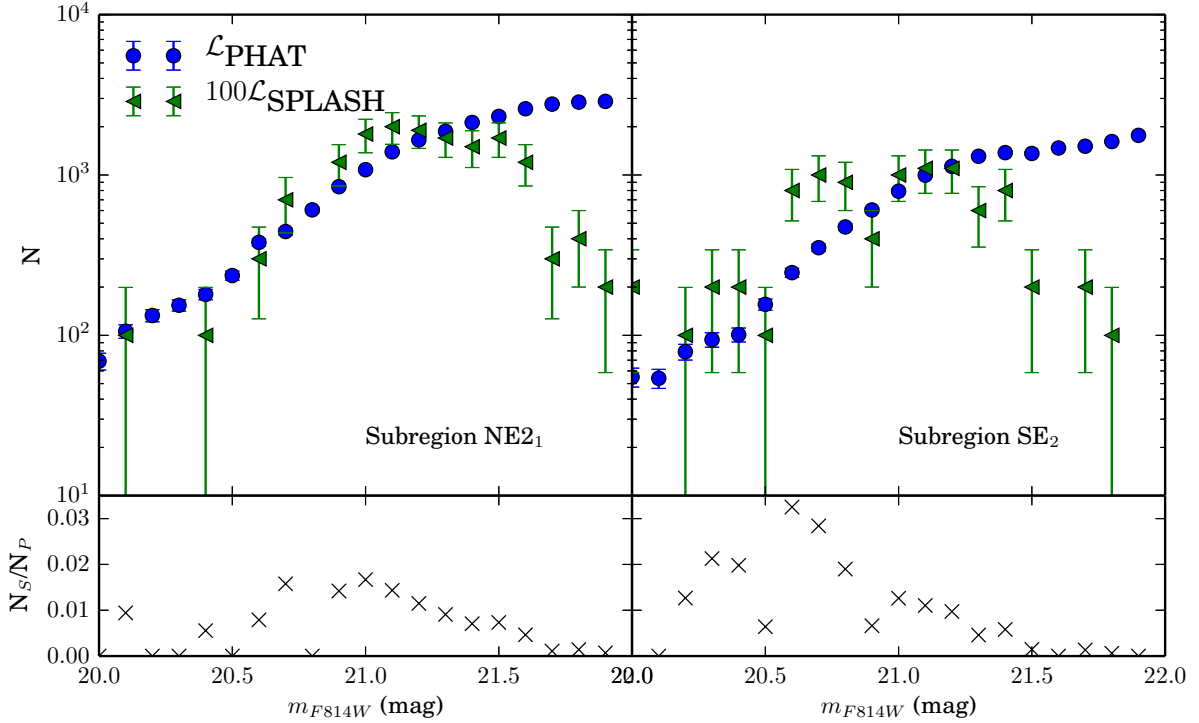}}
\centering
\caption{{\em Top panels: } PHAT (blue) and SPLASH (green) luminosity functions (LFs) in two
  subregions: NE2$_1$, which lies on the major axis, and SE$_2$, which
lies on the minor axis. The SPLASH LF has been scaled by a factor of
100. Error bars represent Poisson uncertainties. Note that the slope
of the PHAT LF changes twice: at $m_{\rm F814W}\sim 20.4$ mag (the
TRGB of a population with $[M/H]\lesssim -0.6$) and around $21$ mag (the
TRGB of a more metal-rich population). {\em Bottom panels: } The
empirical selection function $N_{\rm SPLASH}/N_{\rm PHAT}$ is shown for each
subregion. The selection function varies from subregion to subregion. }
\label{fig_obslf}
\end{figure*}

We display the PHAT and SPLASH luminosity functions from two
representative subregions in Figure~\ref{fig_obslf}. The shape of the
PHAT LF is similar in all subregions: the slope is shallow at the
faint end, gradually steepens brightward of $m_{\rm F814W} \sim
21$, and finally flattens again brightward of $m_{\rm F814W} \sim
20.5$. These changes in slope can be explained by the position of the
TRGB in M31 and the presence of intermediate-age stars brighter
than the TRGB. The magnitude of the TRGB at the
distance of M31 is $m_I = 20.35$ near $[M/H]=-1$, changing by only about
$0.1$ dex between $-2.5 < [M/H] < -0.57$ \citep{sal97}. However, the
metallicity distribution of M31's disk and bulge extend to supersolar
values \citep{sar05, bro06}. At such high metallicities, line
blanketing in the red can push the TRGB even fainter than $m_I =
21$. The wide shoulder seen in the PHAT LF between the TRGB and
$m_{\rm F814W}\sim 21$ is a signature of a broad metallicity
distribution with a broad range of TRGB magnitudes. The shallow slope
brightward of the metal-poor TRGB at $m_{\rm F814W} = 20.35$ suggests a high
fraction of young or intermediate-age populations with bright AGB
stars, as described by \citet{men02}. 

The SPLASH luminosity function varies from subregion to subregion, but
usually peaks in the range $20.5 < m_{\rm F814W} < 21.5$ due to the
design of the spectroscopic survey. (Targets in this magnitude range
were bright enough to yield high-quality spectra, but faint enough to
likely be M31 members rather than foreground MW dwarfs, so they were
given highest priority in the target selection process.) Because there
are so many fewer stars in the SPLASH sample than in PHAT, for display
purposes we have scaled the SPLASH LF by a factor of 100, and scaled
the Poisson errors accordingly.

We can measure the spectroscopic selection function $S_s(M)$ in each
subregion $s$: 

\begin{eqnarray}\displaystyle
  S_s(M) = \mathcal{L}_{\rm SPLASH,s}/\mathcal{L}_{\rm PHAT,s}
\end{eqnarray}

\noindent $S_s(M)$ is a purely empirical measure of the complicated target
selection criteria of the spectroscopic survey. Examples of $S_s(M)$
in two representative subregions are shown in the bottom panels of
Figure~\ref{fig_obslf}. The variation in the selection function with
subregion and with magnitude means that the SPLASH survey may over-
or under-sample the spheroid relative to the disk. We correct for this
effect, as described late in the following section. 

\section{Analysis}\label{sec_analysis}

Our goal is to find the most probable combination of structural and
LF parameters for a S\'ersic bulge,
exponential disk, and power-law halo given the three sets of constraints
described above: surface brightness, PHAT LF in each of 14 subregions,
and kinematically-derived disk fraction in each subregion. As shown in
Figure~\ref{fig_flowchart}, we build a toy model to represent the SB,
disk fraction and LF across the galaxy. In
\S\,\ref{ssec_profiles} we describe our model: a 2D surface brightness
profile and LF for each structural subcomponent. In
\S\,\ref{ssec_likelihood} we show the likelihood function to be
sampled.  Finally, in \S\ \ref{ssec_mcmc}, we describe the Markov
chain Monte Carlo sampler {\tt   emcee} \citep{for12} that we use to
sample the parameter space.

\subsection{Model}\label{ssec_profiles}

The model parameters are listed in the first column of
Table~\ref{tab_params}. All parameters have flat priors within the
range specified in the fourth column of the table. The only fixed
parameter is the bright-end slope (log $N$/mag = $500$/mag) of the halo
LF, chosen for reasons described in \S\,\ref{sss_modellf}. 

\subsubsection{Surface Brightness Profiles}\label{sssection_profiles}

As discussed in the introduction, we choose simple, standard SB
profiles for three components: the bulge, disk, and halo. 
Each profile is given in terms of position on the sky $(R, PA)$, where $R$ is the
projected radius and and $PA$ is the position angle measured east of
north. Hence, we can fit to an entire SB map, rather than only to data
points that happen to fall on the major or minor axis. We assume that
the three model components have the same major axis position angle
$pa$. 

\paragraph{Bulge} For the bulge,
we assume a generalized S\'ersic profile with S\'ersic index $n_b$,
half-light radius $R_b$ and intensity $I_b$ at $R_b$: 

\begin{eqnarray}\displaystyle
  \Sigma_b(R_{\rm eff,b}) = I_b\exp\left\{-A_{n_b}\left[\bigg(\frac{R_{\rm
          eff,b}}{R_b}\bigg)^{1/n_b} - 1\right]\right\}
\label{eq:sersic}
\end{eqnarray}

\begin{eqnarray}\displaystyle
  R_{\rm eff,b}(R, \Delta PA, \epsilon_b) = R\sqrt{\cos^2\Delta \rm PA +
    \bigg(\frac{\sin\Delta \rm PA}{1
      - \epsilon_b}\bigg)^2}
 \label{eq:deproj}
\end{eqnarray}

\begin{eqnarray}\displaystyle
  \Delta PA = PA - pa
\label{eq:dpa}
\end{eqnarray}
%

\noindent where $A_{n_b} =
1.9992n_b-0.3271$ \citep{cap89}. The formula is given in terms of the
effective major axis  (deprojected) coordinate $R_{\rm eff,b}$,
which is a function of the major axis position angle $pa$ and
bulge ellipticity $\epsilon_b$. With $n_b = 1$ or $n_b = 4$, the profile
reduces to the exponential or de Vaucouleurs profile, respectively. 

\paragraph{Disk} We assume an exponential SB profile for the disk: 
\begin{eqnarray}\displaystyle
  \Sigma_d(R_{\rm eff, d}) = I_d\exp{(-R_{\rm eff,d}/R_d)},
  \label{eq:exponential}
\end{eqnarray}

\noindent where $I_d$ is the disk surface brightness at the galactic center and $R_d$ is the 
scale length in the plane of the disk. The formula is given in terms
of deprojected radius $R_{\rm eff,d}$, computed as before in terms of
($R, \Delta
\rm PA$)  and disk ellipticity $\epsilon_d$.

\paragraph{Halo} Finally, we assume a 2D cored power-law
halo SB profile. Though it is also possible to model a halo as a S\'ersic
function, we adopt a power-law because \citet{cou11} demonstrate that
such a model is a more reasonable description of the M31 halo. The
halo surface brightness is then given as

\begin{eqnarray}\displaystyle
  \Sigma_h(R_{\rm eff,h}) = \frac{I_h}{(1 + (R_{\rm eff,h}/R_h)^2)^{\alpha/2}}
  \label{eq:halo}
\end{eqnarray}

\noindent where $I_h$ is the halo surface brightness at the
galactic center, $R_h$ is the radius of the core, and the effective major
axis coordinate $R_{\rm eff,h}$ is
defined as before in terms of coordinates $(R, \Delta PA)$
and halo ellipticity $\epsilon_h$.

We define 

\begin{eqnarray}\displaystyle
  \Sigma_T(R, \Delta PA) = \Sigma_b + \Sigma_d + \Sigma_h
  \label{eq:model}
\end{eqnarray}

\noindent to be the total surface brightness of the model at coordinates $(R,
\Delta PA)$. 

For each profile, we fit for the central intensity in units of $\rm
mag~arcsec^{-2}$ rather than $\rm counts~arcsec^{-2}$. So our model
parameters describing central magnitudes are $\mu_b,~\mu_d,~\mu_h$
rather than $I_b,~I_d,~I_h$. The three central magnitude parameters $\mu_k$
are defined as follows:  

\begin{eqnarray}\displaystyle
\mu_b = -2.5\log_{10} I_b + 25.6 \\
\mu_d = -2.5\log_{10} I_d +25.6 \\
\mu_h = -2.5\log_{10} I_h + 25.6 
\end{eqnarray}

\noindent Here the zeropoint $25.6$ is chosen to match that of the
\citet{cho02} SB data. 

Because the fractional photometric uncertainties $e_m$ on the
SB measurements are typically very
small, we introduce an uncertainty parameter $\epsilon_{\rm SB}$ into
the model to allow for additional photometric uncertainty as well as
departures from the assumed functional form of the SB
profile. $\epsilon_{\rm SB}$ is allowed to have a unique value for
each SB dataset $j$, but may not vary between points $i$
within a given data set. The total fractional uncertainty of a given
SB measurement $\Sigma_{ij}$ is then the quadrature sum of the Poisson
uncertainty and the error parameter: 

\begin{eqnarray}\displaystyle
\epsilon_{ij} = \Sigma_{ij}\sqrt{e^2_{m,ij} + \epsilon_{{\rm SB},
    j}^2}
\end{eqnarray}

\subsubsection{Disk Fractions}\label{sssection_modeldiskf}

We have measured the disk fraction distribution $\phi_s(f_d)$ in each
subregion $s$, and we want to know what disk fraction our SB model
predicts. The fraction of integrated I-band light contributed by the
disk is of course $\Sigma_d/\Sigma_T$. However, to compare this to a
kinematically-derived disk fraction, we must convert it to a fraction
of stars, as sampled by the somewhat arbitrary SPLASH survey,
contributed by the disk. This conversion requires knowledge of the
intrinsic disk luminosity function $\mathcal{L}_d$, the total
luminosity function $\mathcal{L}_T$, and the subregion-specific SPLASH
selection function $S_s(m_{\rm F814W})$. The model disk fraction in
SPLASH star count units in subregion $s$ is

\begin{eqnarray}\displaystyle
  f_{\rm d,s} = \frac{\int  S_s n_{d,s} \mathcal{L}_d
    dm_{\rm F814W}}{\int S_sn_{t,s}\mathcal{L}_{T,s}dm_{\rm F814W}}
  \label{eq:diskf}
\end{eqnarray}

\noindent where the integration only needs to be performed over the magnitude
range $m_{\rm F814W}=[20, 22]$, where the SPLASH selection function is
nonzero. 

Later in this paper (section \ref{ssec_c}), we will use a conversion
factor $C_s$ to convert between disk fraction units. $C_s$ is
defined as the ratio between $f_{\rm d,s}$ (the disk fraction as
measured in SPLASH star counts, as defined in Equation~\ref{eq:diskf})
and the disk fraction as measured in SB units: 

\begin{eqnarray}\displaystyle
  C_s \equiv \frac{ f_{\rm d,s}}{\Sigma_{d,s}/\Sigma_{T,s}}.
  \label{eq:Cs}
\end{eqnarray}

$C_s$ is constant within a subregion $s$, but varies from subregion to
subregion. 

As with the SB data, we introduce a kinematical uncertainty
parameter $e_k$ (expressed as a fraction of the empirical skew-normal
width $\sigma_s$) to account for non-ideal $f_{\rm d}$ calculation due
to, e.g., any non-Gaussianity in the disk line-of-sight velocity
distribution. We add this parameter in quadrature to $\sigma_s$ and
re-normalize the widenened skew-normal PDF. 

\subsubsection{Luminosity Functions}\label{sss_modellf}

Equation~\ref{eq:diskf} shows that computing the predicted disk
fraction $f_{\rm d}$ (in star count units) requires knowledge of the
disk luminosity function $\mathcal{L}_d$ in the magnitude range
$m_{\rm F814W}=[20, 22]$. Therefore, we model the bright end of the
disk, bulge, and halo luminosity functions. This magnitude range
includes the tip of the
red giant branch (TRGB), which lies near $m_{\rm F814W}=20.5$ at the
distance of M31 for stars with $[M/H] \lesssim -0.5$ and fainter for more
metal-rich populations (see discussion in \S\,\ref{ssec_lf}). 

Because RGB stars dominate the stellar population faintward of the
TRGB, while younger objects such as AGB stars fill in the brightward
portion of the LF, there is reason to expect that the number density
of stars should fall at different rates faintward and brightward of
the TRGB. Hence, we parameterize the LF of each component $k$ = (disk,
bulge, halo) as a broken powerlaw in log(number density) vs. magnitude
space: 

\begin{eqnarray}\displaystyle
  \log\mathcal{L}_k(m_{\rm F814W}) = c_{0k} +  
    \begin{cases}
      p_km & {~m < l_k} \\
      q_km+ l_k(p_k - q_k) & {~m\ge l_k} \\
      \end{cases}
 \
  \label{eq:model}
\end{eqnarray}

\noindent where $c_{0k}$ is chosen such
that $\mathcal{L}_k$ is normalized to unity over this magnitude
range. 

Since it is
reasonable to expect that the stellar disk may have an age or
metallicity gradient, we allow each of the disk LF parameters to depend
linearly on radius on the plane of the disk $R_{\rm  eff,d}$ in the
radial range of interest $5~{\rm kpc}< R_{\rm eff,d} < 20$ kpc: 

\begin{eqnarray}\displaystyle
    p_d = p_{d0} + \delta p_d \ln(R_{\rm eff,d})\\
    q_d = q_{d0} + \delta q_d \ln(R_{\rm eff,d})\\
    l_d = l_{d0} + \delta l_d \ln(R_{\rm eff,d})
    \label{eq:p-gradient}
\end{eqnarray}


However, for simplicity, we require that the power-law slopes $p, q$
and the break magnitude $l$ for the bulge and the halo be constant
with radius. This assumption does not affect our results, since the
portion of the galaxy covered by the LF and kinematical surveys is
almost entirely disk-dominated according to SB-only decompositions
\citep{cou11} as well as our decomposition. Even if the halo does
contribute a significant number of stars in the SPLASH survey region,
its range of metallicities is confined to the metal-poor regime
\citep{kal06}, where the magnitude of the TRGB is insensitive to
metallicity. Near the end of Section 6.1, we discuss this point
further. 

We require for each component that $p, q$ be nonnegative within the
radial range $5~{\rm kpc}< R_{\rm eff,d} < 20$ kpc,
and that $l$ lie within the magnitude range of  interest,
$m_{F814W}=[20,22]$. We also require that the bright-end slope of the
halo LF to be extremely steep ($p_h = 500$, because an old component
should not have stars brighter than the TRGB) and that the break magnitude
of each component be fainter than the brightest expected TRGB
components ($l_k < 20.3$). 

The total normalized LF ${\mathcal L_{T,s}}$ in subregion $s$ is then the
weighted sum of those of the disk, bulge, and halo, where the weights
correspond to the number density of stars in each subcomponent in the
magnitude range of interest. To compute the weights, we assume that in
each component, the ratio of the number density $n_k$ of  stars in the
range $I= [20,22]$ to the surface brightness $\Sigma_k$ of that
component is a constant $y_k$. Hence, we introduce four new model
parameters $y_b, ~y_h,~y_{d0},$ and $\delta y_d$, defined by the
following relationships:

\begin{eqnarray}\displaystyle
  y_b \equiv \frac{n_{b,s}}{\Sigma_{b,s}}\\
  y_h \equiv \frac{n_{h,s}}{\Sigma_{h,s}}\\
  y_d(R_{\rm eff,d}) \equiv \frac{n_{d,s}}{\Sigma_{d,s}}\\
  y_d(R_{\rm eff, d}) = y_{d0} + R_{\rm eff, d}\delta y_d
  \label{eq:y_i}
\end{eqnarray}

\noindent where $\Sigma_{k,s}$ is the average surface brightness of
subcomponent $k$ integrated over the area of subregion $s$. The ratio $y_k$ is a
constant independent of subregion $s$ for the bulge and
for the halo, and depends linearly on $R_{\rm eff,d}$ for the disk. Then the total
number density of stars in subregion $s$ is 

\begin{eqnarray}\displaystyle
n_{t,s} = (y_{d0} + R_{\rm eff,d}\delta y_d)\Sigma_{d,s} + y_b\Sigma_{b,s} + y_h\Sigma_{h,s} 
\label{eq: n_T}
\end{eqnarray}

\noindent and the normalized model LF is 

\begin{eqnarray}\displaystyle
  {\mathcal L}_{T,s}(m)=\frac{n_{d,s}{\mathcal L}_d +
  n_{b,s}{\mathcal L}_b + n_{h,s}{\mathcal L}_h}{n_{t,s}}
  \label{eq:lftotal}
\end{eqnarray}

As with the SB and
kinematical data, we introduce an uncertainty parameter
$\epsilon_{\mathcal{L}}$ to account for differences between the
assumed broken power law and the actual shape of the
LF. $\epsilon_{\mathcal{L}}$ is a fractional uncertainty whose value
is constant across all magnitude bins and all subregions. The
total fractional uncertainty on a given LF bin $m$ and subregion
$s$ is the quadrature sum of the Poisson fractional uncertainty and
$\epsilon_{\mathcal{L}}$. Then the total uncertainty on that LF bin
is  

\begin{eqnarray}\displaystyle
  \epsilon_{LF,m,s} = \mathcal{L}_{\rm PHAT,m,s}\sqrt{\frac{1}{N_{m,s}} + \epsilon_{\mathcal{L}}^2}.
\end{eqnarray}

\subsection{Model-Data Comparison: Likelihood
  Function}\label{ssec_likelihood}
The probability that a point in parameter space is a good
representation of the data is given by the product of three
goodness-of-fit statistics describing the agreement between the model
and the three data sets: 

\begin{eqnarray}\displaystyle
P = P_{\rm SB}P_{\rm LF}P_{\rm f_d}
\end{eqnarray}

We work instead with the log-likelihood: 

\begin{eqnarray}\displaystyle
    \ln P = \ln P_{\rm SB} + \ln P_{\rm LF} + \ln P_{\rm f_d}
    \label{eq:likelihood}
\end{eqnarray}

The SB factor is summed over each data point $i$ in each data set $j$:

\begin{eqnarray}\displaystyle
  \ln P_{\rm SB} = -\sum_{i,j = 1}^{N,n_i}\left[\frac{(\Sigma_{T,ij} -
    \Sigma_{{\rm obs},ij})^2}{2\pi \epsilon_{ij}^2}\right]
  \label{eq:L1}
\end{eqnarray}

Meanwhile, the LF and disk fraction factors are
summed over each of the 14 subregions. The
goodness-of-fit to the disk fraction is the
height of the skew-normal function $\phi_s(f_{\rm d})$ that describes
the probability distribution of the kinematically
measured disk fraction in subregion $s$, evaluated at the model disk fraction
$C_s\frac{\Sigma_{d,s}}{\Sigma_{T,s}}$:

\begin{eqnarray}\displaystyle
 \ln P_{\rm f}= \sum_{s = 1}^{14}\ln \phi_{s}\bigg(f_{\rm d} = C_s\frac{\Sigma_{d,s}}{\Sigma_{T,s}}\bigg)
  \label{eq:L2}
\end{eqnarray}

The LF component of the likelihood is determined by
summing the difference between the observed and model total luminosity
functions over every magnitude bin $m$ in every subregion $s$:

\begin{eqnarray}\displaystyle
 \ln P_{\rm \mathcal{L}} = -\sum_{s,m=
   1}^{14,20}\frac{(n_{p,s}\mathcal{L}_{{\rm PHAT},s,m} -
    n_{T,s}\mathcal{L}_{T,s,m})^2}{\epsilon_{LF,  s, m}^2}.
  \label{eq:L3}
\end{eqnarray}

\subsection{MCMC Sampler}\label{ssec_mcmc}
To estimate the probability distribution function of each model
parameter, we draw samples from the log-likelihood function
(Equation~\ref{eq:likelihood}) using the
the Markov chain Monte Carlo sampler {\tt emcee} \citep{for12}. More
details on {\tt emcee} can be found in the Appendix. This section will
only summarize the details unique to this paper. 

In {\emcee}, and more generally MCMC, an ensemble of ``walkers'' --- or
points in parameter space --- moves
through parameter space. During each step, each walker is given the
option to move a specified distance along the line in parameter space
connecting it to a random other walker. Moves corresponding to
increases in the value of the likelihood function are always accepted;
moves corresponding to decreases in likelihood are sometimes
accepted. After many steps (the ``burn-in'' phase), the distribution
of walkers samples the likelihood function: the density of walkers is
highest in high-probability regions of parameter space. In this paper,
we allow 256 walkers to burn in for 10,000 steps, and analyze the
probability distributions using their positions in their last 100
steps. These probability distributions are shown in
Figures~\ref{fig_dfa} and \ref{fig_dfb} in the Appendix. 

\section{Results}\label{sec_results}

\subsection{Confidence Intervals \& Correlations} 
We measure the mean value and $1\sigma$ confidence interval for each
parameter from its posterior probability distribution. We report these statistics
in Table~\ref{tab_params}. In general, the parameters describing the
SB profiles are very well constrained, to better than $1\%$, while the
LF parameters are constrained to $10\%$ at best. 

\begin{figure*}[h]
\scalebox{1.0}{\includegraphics[trim=30 150 30 70, clip =
  True]{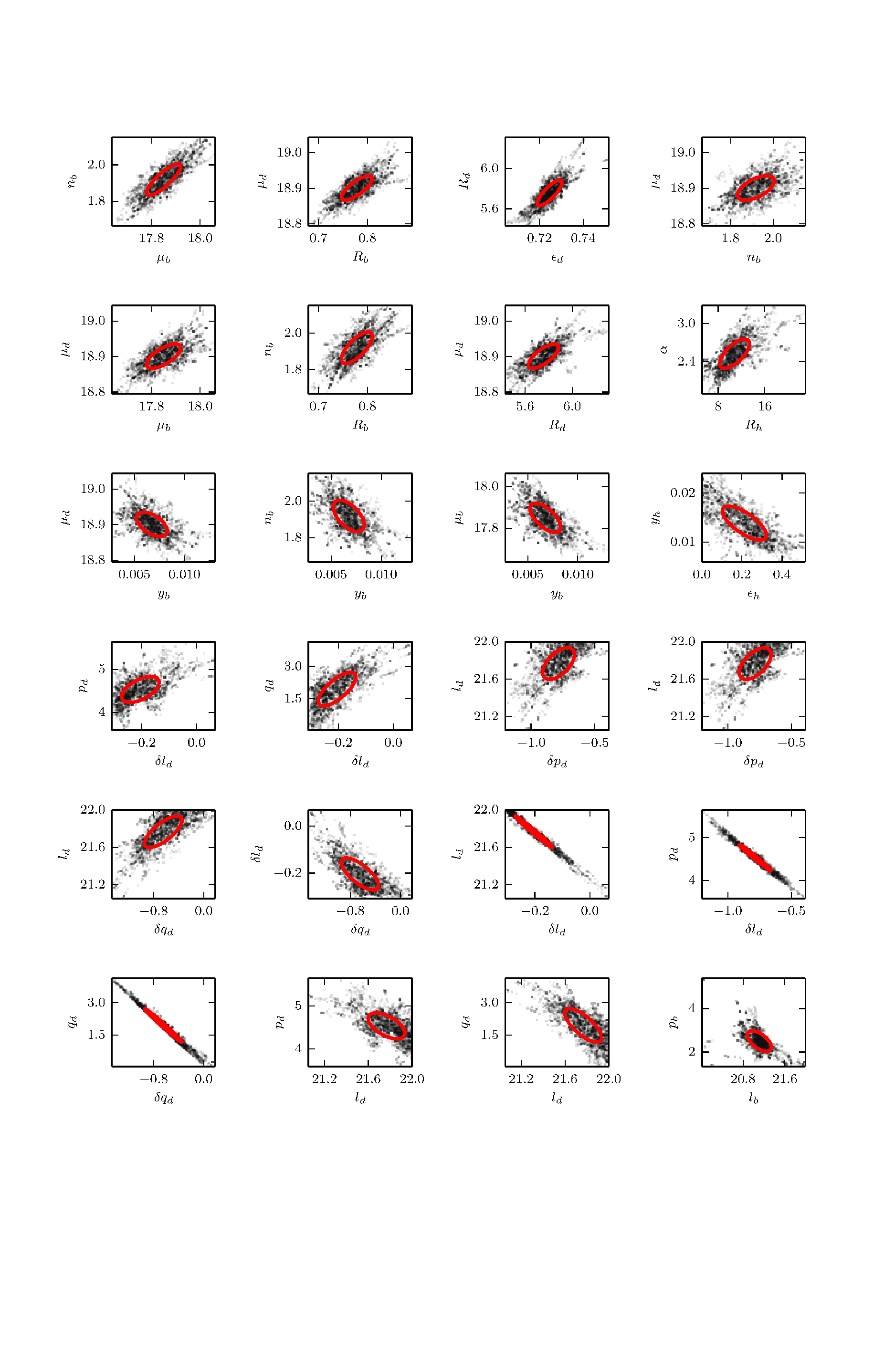}}
\centering
\caption{2D posterior distributions of 24 of the strongly (Spearman
  $r^2>0.6$) or significantly ($0.6\geq r^2>0.25$)
  correlated pairs of parameters. (Only 28 of the 512 pairs of
  parameters fall into one of these categories.) Red dashed contours
  are $1\sigma$ error ellipses. In general, a parameter is correlated
  only with other parameters describing the same subcomponent (bulge,
  disk or halo). LF parameters tend to be correlated only with other LF
  parameters, and SB parameters only with other SB parameters. } 
\label{fig_corr}
\end{figure*}

Some of the parameters appear in Table~\ref{tab_params} to be
completely unconstrained, but in fact are simply strongly correlated
with other parameters. For a given such pair, the
allowed region in 2D space can be quite small. To quantify covariances
between parameters, we calculate the Spearman correlation coefficient
$r$ between every pair of parameters. Those pairs with $r^2 > 0.6$ are labeled
``strongly correlated'', while those with $0.25 < r^2 < 0.6$ are labeled
``significantly correlated.''
Figure~\ref{fig_corr} shows the 2D probability distributions of 24 of
the 28 strongly or significantly correlated pair of parameters.

All six of the strongly correlated pairs,
and all but four of the significantly correlated pairs, consist
of two parameters describing the same subcomponent. For example, the 
bulge effective radius $R_b$ is significantly correlated with the bulge
S\'ersic index $n_b$, but not with the disk or halo scale
radii. Similarly, the disk LF bright-end slope $p_{d0}$ depends on the
disk LF break magnitude $l_{d0}$, but not on the slopes or break
magnitudes of the bulge or halo LFs. Additionally, with a few
exceptions, the SB parameters tend to be correlated only with other SB
parameters, while the LF parameters tend to be correlated only with
other LF parameters. 

\subsection{Quality of Profile Fits}
\begin{figure}
\scalebox{0.63}{\includegraphics[trim=0 100 10 135, clip =
  true]{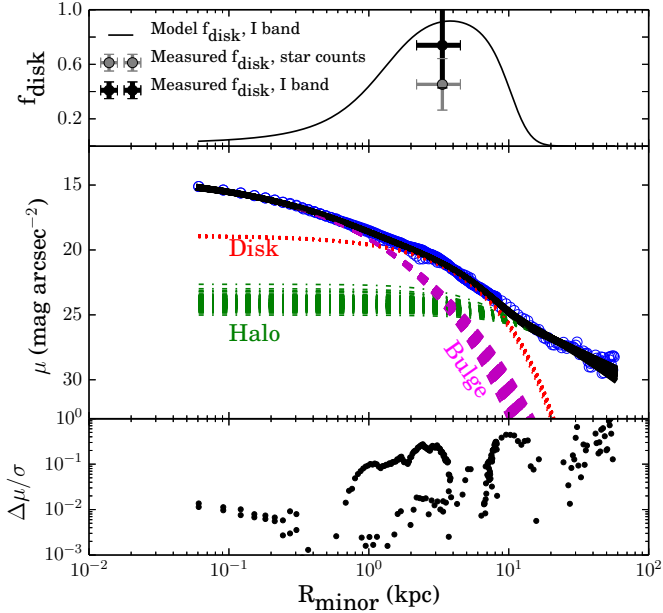}}
\centering
\caption{Minor axis SB decomposition. {\em Middle:} Fit of the model
  SB profile (black lines) to the observed SB profile (blue circles) as a function
  of projected radius. The model is
  also shown decomposed into the disk (red dotted), halo (green dot-dashed), and bulge
  (magenta dashed) components. One line is displayed for each of 256 samples
  drawn from the posterior distribution of walkers, so that the width
  of each region encloses the entire uncertainty (not the $1\sigma$
  uncertainty) associated with that profile. The disk, inner bulge, and outer halo
profiles are well constrained, whereas the inner halo and outer bulge
profiles are less well constrained. {\em Top:} Model disk
fraction in SB units (black line) slightly overpredicts kinematically measured disk
fraction in the minor axis subregion ($SE_2$) converted to SB units (black
cross). The conversion factor from disk fraction in star counts (gray
cross) to disk fraction in SB units (black cross) is about $1.2$ in
this subregion: the kinematical survey oversamples the spheroid population. {\em
  Bottom}: Average residual between model and measured SB as a
function of projected radius, relative to the measurement uncertainties on
the SB. The lower and upper tracks between 0.3 and 2.5 kpc
correspond to the Choi and Irwin data sets, respectively.} 
\label{fig_sbp_sersic}
\end{figure}

Because the degeneracy between {\em parameters} is generally confined to
within a subcomponent, the subcomponent SB and LF {\em profiles} are 
well constrained and are not degenerate with one another. The middle panel of
Figure~\ref{fig_sbp_sersic} shows a minor axis projection of the SB
decomposition into bulge,
disk, and halo subcomponents. One set of
profiles (bulge, disk, and halo) is displayed for each of 256 samples
drawn from the last step of the walker ensemble. The set of 256 violet
lines, then, samples the entire range of bulge profiles allowed by
the data; similarly, the set of 256 red lines represents the
entire allowed range of disk profiles. Note that the profiles are
relatively well constrained: for example, while the bulge scale radius
$R_b$ and S\'ersic index $n_b$ each vary significantly, they covary in
such a way that the bulge is always small, contributing less SB
than the halo at only $R\sim4$~kpc on the minor axis. 

\begin{figure}[h!] 
\scalebox{1}{\includegraphics[trim=20 185 100 65, clip =
  true]{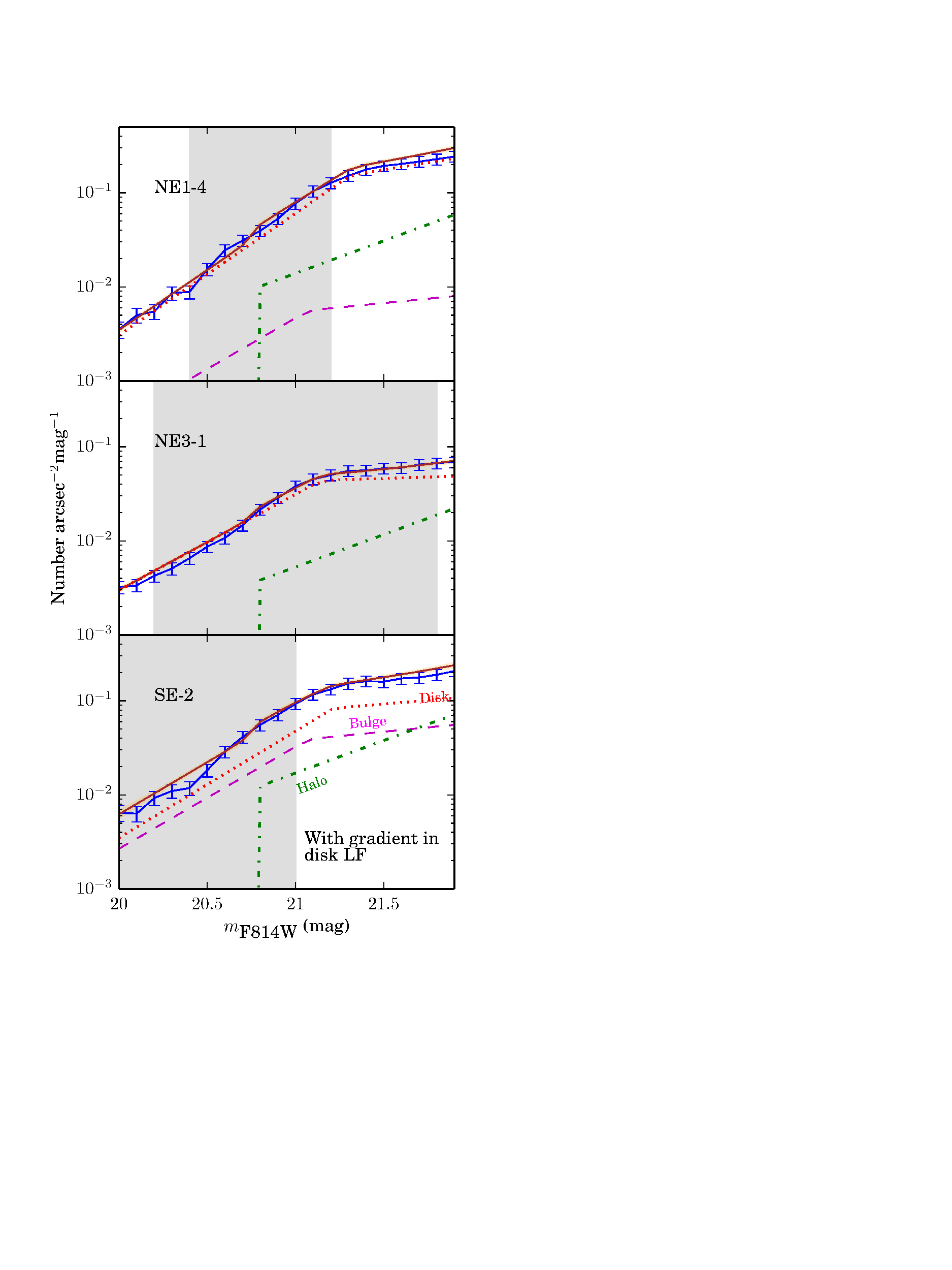}}
\centering
\caption{Comparison between observed PHAT (blue) and model (brown)
 total normalized LFs after a fit to the SB with
 kinematical constraints, assuming radial gradients in the disk LF
 shape parameters, in three representative subregions. The shaded portion shows
 where the SPLASH selection function is at least 30\% of its maximum
 value. The model is shown decomposed
into disk (red), bulge (magenta), and halo (green) components. Each LF is
weighted by the number density of stars in the magnitude range $F814W
= [20, 22]$ in that subcomponent in the subregion of interest. The
width of the brown region is given by the variation in the model
LF. Error bars on the PHAT LF include contributions from both the $\sqrt{N}$
Poisson uncertainty in each $0.1$ mag bin and the LF uncertainty
parameter. }
\label{fig_lferr}
\end{figure}

\begin{figure}[h!]
\scalebox{0.7}{\includegraphics[trim=0 0 30 30, clip =
  true]{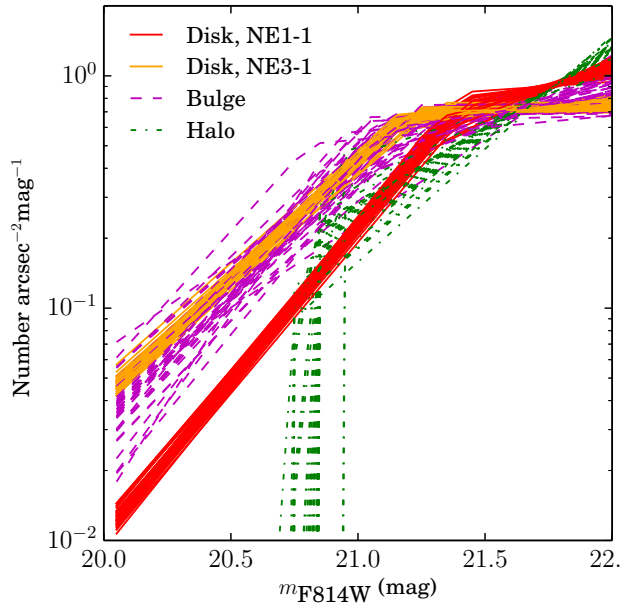}}
\centering
\caption{Normalized median bulge (magenta), disk (red, orange), and halo (green)
  LFs. For each component, one line is displayed for each of 256 samples
  drawn from the posterior distribution of walkers, so that the width
  of each region encloses the entire uncertainty (not the $1\sigma$
  uncertainty) associated with that profile. The disk LF depends on
  radius in the plane of the disk $R_{\rm d}$; the disk LFs in the
  subregions with smallest and largest $R_{\rm d}$ are displayed in
red and orange, respectively, to illustrate the range in disk LF over
the PHAT survey region. }
\label{fig_normlf}
\end{figure}

The LF decomposition in each of three representative subregions is
shown in Figure~\ref{fig_lferr}. The disk (red dotted line) dominates
in every subregion, so that it is nearly indistinguishable from the
total model LF (orange shaded region) or observed PHAT LF (blue line
with error bars). For clarity, in this plot we only display the bulge, disk, and
halo LFs corresponding to the median values of the
parameters. However, in Figure~\ref{fig_normlf} we display the three
LFs normalized in the magnitude range $20 < m_{\rm F814W} < 22$, with
one line drawn for each of 256 samples in the walker ensemble. Since
the shape of the disk LF is allowed to change with radius, we display
two representative disk LFs: one at the radius of an inner subregion
(red lines) and one at the radius of an outer subregion (gold
lines). The disk LF is tightly constrained at any given radius despite
the significant degeneracy between the slopes and central values of
the individual LF parameters. The bulge (violet) and halo
(green)  LFs are also relatively well constrained, especially
considering that neither component dominates in any subregion sampled
by our portion of the PHAT survey. 

\begin{figure}[h!]
\scalebox{0.6}{\includegraphics[trim=80 0 95 35, clip = true]{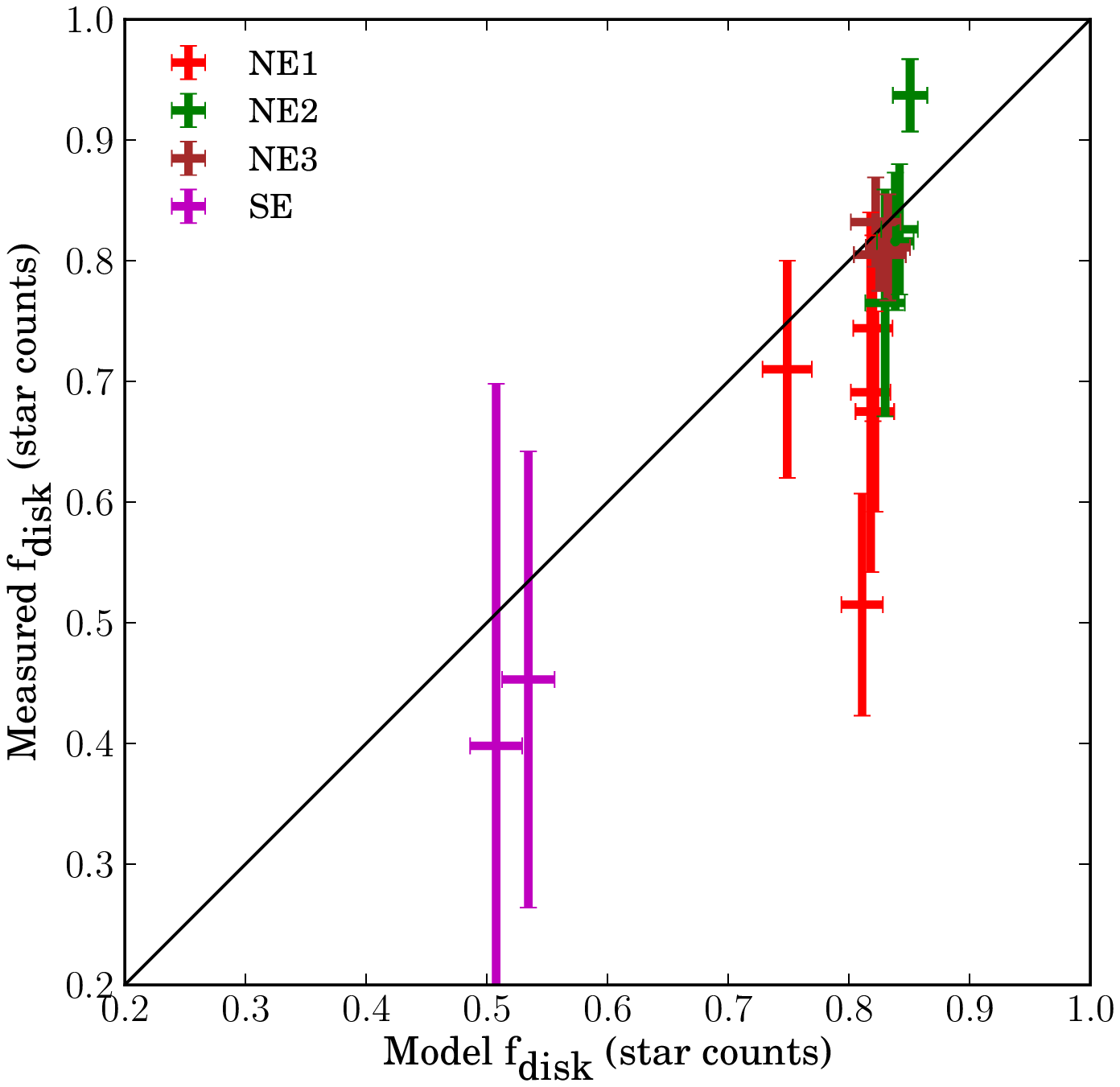}}
\centering
\caption{Comparison between the
  measured cold and model disk fractions, in units of SPLASH star counts, in the final
  decomposition. Error bars reflect estimated $1\sigma$ uncertainties
  from the appropriate distributions at the end of the MCMC
  chain. Points are color-coded by spatial region. Measured cold
  fractions are systematically lower than model disk fractions,
  suggesting that there may be a dynamically hot (``kicked-up'') disk
  component for which our model has not accounted.}
\label{fig_fracline_sersic}
\end{figure}

The model fits the SB and LF data sets quite well. The bottom panel of
Figure~\ref{fig_sbp_sersic} displays the
difference between the observed and best-fit SB profiles expressed in units
of SB uncertainty (a combination of Poisson measurement errors and
error parameter). The fit to the SB is very good: the magnitude error
is generally less than 10\% of the SB uncertainty, corresponding to a
median reduced $\chi^2$ of $0.82$. The fit to the LF is also
acceptable, with a median $\chi^2$ of $0.81$. However, the fit to the
kinematics is less satisfactory (median reduced
$\chi^2=1.88$).  Figure~\ref{fig_fracline_sersic} shows that the model
overestimates the disk fraction in almost every subregion, and by more
than $1\sigma$ in four subregions. Though the measured disk fractions
do constrain the model, their limited number and relatively large
uncertainties mean that they have less effect on the final
decomposition than the LF and SB data, which strongly prefer a
small S\'ersic component and dominant exponential component. The poor
fit to some of the
kinematical measurements is then a sign of tension between our (very
simple) model and the data. In \S\,\ref{ssec_diskf}, we show that
allowing for a dynamically hot ($\sigma_v \sim 150~\rm km~s^{-1}$)
component with disklike population and spatial profile (a ``kicked-up
disk'') can reduce this tension. 

The decompositions presented in Figures~\ref{fig_sbp_sersic} and
\ref{fig_normlf} will be discussed in more detail later in this
section (\S\,\ref{ssec_previous}) and in the Discussion
(\S\,\ref{sec_discussion}). 

\subsection{Comparison to previous measurements}\label{ssec_previous}

The median values of our parameters are presented in
Table~\ref{tab_params}. Here, we compare some of the structural
parameters to previous measurements. 

The disk scale length we measure, $5.76\pm0.1$ kpc, is consistent with the
range of accepted values. \citet{wor05} measure an
$R_d=5.8\pm0.2~\rm kpc$ from elliptical isophotal fits to the Choi
I-band image alone. \citet{sei08} measure $R_d =
5.91\pm0.27$ kpc from an IRAC $3.6~\mu m$ profile. \citet{iba05}
measured a scale length of $5.1\pm0.1~\rm kpc$ by fitting an
exponential profile to RGB star counts in the range $R =
20-40$ kpc on the major axis --- though if the spheroid actually
contributes some additional light in the inner part of that radial range,
the disk profile could appear to fall off faster than it actually
does, and so the authors would underestimate the disk scale length. 

Our $R_d$ is nearly $0.8$ kpc ($16\%$) larger than that measured in
the same band with an SB-only decompostion (Model U in
\citet{cou11}). This offset is likely to arise because we fit for
the position angle of the disk, whereas \citet{cou11}
simply assume a position angle of $37.7^\circ$. The scale length of a
disk is largest when measured along its true major axis (that is, when
the position angle is correct). 

Our median disk ellipticity, $0.725\pm0.005$, corresponds to an
inclination of $74.0^{\circ}\pm0.3^{\circ}$ in the infinitely thin
limit. The HI disk at similar radii has a similar inclination
\citep{che09,cor10}. Overall, the optical disk appears about three degrees
closer to edge-on, though M31's badly warped outer disk \citep{che09,
  cor10} implies a strong dependence of inclination on
radius. Additionally, the older
populations that we sample are more likely to have been heated and thickened via
satellite interactions at some point in their lifetimes, making the
RGB disk appear more face-on than the younger stellar disk.

Our disk position angle ($44.4^\circ \pm0.5^\circ $) is $6-8$ degrees
greater than that of the HI disk in the $R\sim 5-20$ kpc
range \citep{che09,cor10}, but is similar to
that seen in the azimuthal number density of all RGB stars in
the PHAT dataset \citep{dal12}. 

Our bulge is nearly identical to that found in the
SB-only decomposition of \citet{cou11}. The S\'ersic index
($1.92\pm0.08$) is intermediate between 
exponential and de Vaucouleurs, characteristic of a disky
``pseudobulge.'' \citep{kor04}. 

\citet{ath06} and \citet{bea07} have found that NIR photometry and
HI/H$\alpha$ kinematics are reproduced by N-body models with at least
two bulge components: a small classical bulge, a larger boxy
pseudobulge, and a bar that may extend beyond the pseudobulge. We do
not distinguish between these subcomponents; instead, our $n=1.9$ S\'ersic bulge
includes all of them. As described later (end of \S\,\ref{ssec_diskf}),
we also try to model a two-component bulge by fitting the
sum of an exponential, a power-law, and {\em two} S\'ersic profiles to
the composite data set, but the highest-likelihood models are
exclusively the single-S\'ersic ones. 

The core radius of the halo, $R_h=10.6^{+2.5}_{-2.0}$ kpc, is
signficantly larger than the value of $3$ kpc measured from a fit to
the number density profile of resolved blue horizontal branch (BHB) stars
\citep{wil12}. This quantity is in general difficult to measure, since
it relies on tracing the halo in the inner regions of the galaxy,
where the faint halo is strongly subdominant to the  disk and
bulge. Our technique is one of the few that does not involve fixing
other structural parameters or using only a single halo tracer such as
BHB stars or RGB stars. As shown in Figure~\ref{fig_corr}, $R_h$ is
degenerate with the power-law halo slope $\alpha$, but is very unlikely to
be shorter than $8$ kpc and is not degenerate with any other
parameters.

The power-law slope of the halo profile external to the core is
$-2.5\pm0.2$, consistent with measurements from SB-only
decompositions \citep{cou11}, resolved RGB star counts
\citep{guh05,gil12} and BHB star counts \citep{wil12}. A
projected surface density power-law slope of $-2.5$ corresponds
to a deprojected density that scales approximately as $r^{-3.5}$, in
good agreement with cosmologically motivated simulations in which
stellar halos are built up via accretion
\citep[e.g.,][]{bul05,coo10} and accretion plus in situ star formation
\citep[e.g.,][]{fon11}. The outer slope of a \citet{nav96} cold dark
matter halo also scales as $r^{-3.5}$, suggesting that stars may trace the dark matter profile
--- but not significantly affect its shape --- at large radii. 

\subsection{Conversion between integrated-light and star-count disk fractions}\label{ssec_c}
The conversion factor between disk fraction as measured in integrated
light and disk fraction as measured in SPLASH star counts ($C_s$ in
Equation~\ref{eq:Cs})  is $0.84-0.86$ in the subregions near
the major axis and $0.62-0.65$ near the minor axis. $C<1$ implies that
the spectroscopic survey is biased towards spheroid stars --- not
a surprise, since the spectroscopic target selection strategy
prioritizes less crowded objects and spheroid stars are less likely to
be located in clusters or clumpy structures. In the discussion that
follows, as in the analysis, this correction has been applied to the model
so that both the model and measured disk fractions
represent the fraction of stars, as sampled by the SPLASH survey, that
contribute to the disk. 

\section{Discussion}\label{sec_discussion}

We discuss three classes of new results. First, we argue that a fraction
of the stars in M31's dynamically hot inner spheroid may have
originated in the disk.  Second, we discuss the transition between the
bulge and halo. Third,  we discuss the evidence for a radial gradient
in the LF of M31's stellar disk.

\subsection{Kicked-up disk}\label{ssec_diskf}

Despite the significant dynamically hot population in the kinematical
sample, both the SB and LF data are best fit by a model with a bulge
too small to contribute light in the SPLASH survey
region. Figures~\ref{fig_sbp_sersic}  and Figure~\ref{fig_lferr}
illustrate the fits. Figure~\ref{fig_sbp_sersic} shows that the SB profile on the
minor axis between projected radii of $1-10$ kpc --- the radial range covered
by the overlap of the SPLASH and PHAT surveys --- is well fit by an exponential
profile. Figure~\ref{fig_lferr} shows that the observed LF is also
well fit by a decomposition where nearly all of the stars belong to
the disk component. Even without trusting the simple toy model
decomposition, the shape of the observed LF from PHAT looks disklike in every
subregion. The change in slope of the LF at $m_{\rm F814W}\sim 20.5$
corresponds to the TRGB of low- or intermediate-metallicity
populations at the distance of M31. In a region dominated by an old
population, the LF would drop off steeply brightward of the TRGB; only
in a population with a significant young- or intermediate-age
fraction (as expected for a disk) can bright stars such as asymptotic
giant branch (AGB) stars fill in the bright part of the LF and give it
the shallow slope as seen in Figure~\ref{fig_lferr}. 



In summary, the SB and LF data suggest that the region sampled by the
SPLASH and PHAT surveys is completely disk dominated, even though the
kinematics reveal a dynamically hot population in this
region. Figure~\ref{fig_fracline_sersic} illustrates this tension:
even in the best simultaneous fit to the LF, SB, and kinematical data,
the kinematically derived disk fractions are systematically lower than
the model disk fractions. The inability to simultaneously fit the three
data sets is a sign of tension between our (very
simple) model and the composite data set. In this section, we propose
a dynamically hot (``kicked-up'') disk component as a possible
resolution, and walk through some other modifications to the model
that cannot resolve the tension. 

We have assumed that the population with a disk LF is exclusively
dynamically colder than the spheroid, but this may be too
restrictive. The tension between the model and the kinematics can be
explained if the dynamically hot component identified in the
kinematics is inflated by a contribution from stars that were born in
the disk but dynamically heated. This population should have a
disklike LF and follow the disk SB profile, but have
spheroidlike kinematics. Note that such a component is not the same as
a thick disk. The velocity dispersion of M31's thick disk is only
about $40\%$ larger than that
of the thin disk \citep{col11}. In contrast, a kicked-up disk has a
velocity dispersion similar to that of the halo: $\sim 150~\rm
km~s^{-1}$, or more than $300\%$ larger than that of the thin disk
\citep{dor12}. In our kinematical decompositions, the ``cold''
fraction includes contributions from both the thin and thick disks,
whereas the ``hot'' fraction includes contributions from the spheroid
and the kicked-up disk. Of course, it is possible that thick disks are
created via a heating mechanism similar to (though less extreme than)
that that creates the dynamically hot kicked-up disk. 

Kicked-up disk stars have been seen in cosmological simulations for
some time. N-body and hydrodynamical simulations
predict that minor accretion events can disrupt galactic disks,
kicking disk stars enough so that they would be kinematically
classified as spheroid members, though they retain some of their
angular momentum \citep{pur10,mcc12}. \citet{pur10} find that a merger with
mass ratio $1:10$ can kick $1\%$ of the disk stars into the halo; this
percentage corresponds to a mass similar to the mass accreted from the
incoming satellite itself. \citet{tis13} find that $3-30\%$ of the
halo mass in members of a suite of MW-like galaxies from Aquarius consist of
stars that originated in the disk. There is some observational evidence for
this as well. \citet{she12} identify M giants in the MW halo with
velocities and abundances consistent with a kicked-up disk
origin. M31 is in the process of merging with the progenitor of the
Giant Southern Stream (GSS), which likely was first tidally disrupted
less than 1Gyr ago \citep{far08}, so it is possible that it hosts a
nonvirialized kicked-up disk component. [However, note that the GSS
progenitor was much smaller than 10\% of the mass of M31 \citep{far08}.]

\begin{figure}[h]
\scalebox{0.5}{\includegraphics[trim=00 0 00 00, clip =
  true]{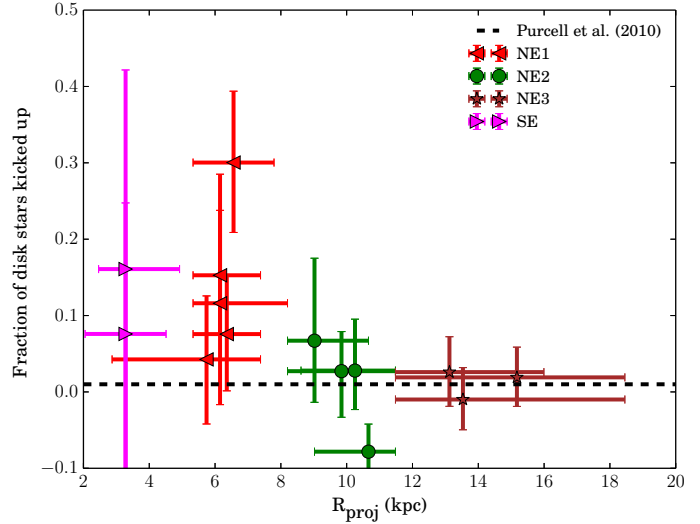}}
\centering
\caption{Fraction of disk-like population that must be dynamically hot
in order to simultaneously fit the LF and kinematics. Vertical error bars
denote $1\sigma$ uncertainties and horizontal error bars denote
the entire range of $R_{\rm proj}$ subtended by each subregion. The kicked up
fraction in most of the subregions is consistent with, though
systematically larger than, that predicted by \citet{pur10} to result
from a merger event at low impact angle with mass ratio 1:10. }
\label{fig_deltaf}
\end{figure}

\begin{figure}[h]
\scalebox{0.5}{\includegraphics[trim=00 0 00 00, clip =
  true]{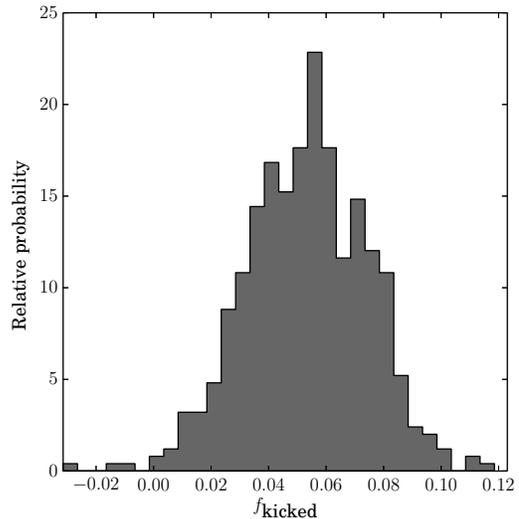}}
\centering
\caption{PDF of $f_{\rm kicked}$, the fraction of disk stars that must be dynamically
  hot in order to simultaneously fit the LF and kinematics. This
  probability distribution is constructed from the 14 PDFs
  corresponding to the individual subregions, weighted by the number
  of stars in each subregion. The most probable value of $f_{\rm
    kicked}$ is 5.2\%. The width of the distribution, a combination of
  spatial variation in $f_{\rm kicked}$ and uncertainty on the mean,
  is $1\sigma$ = $2.1\%$. The
  distribution is narrow enough that a model with zero kicked-up
  fraction is effectively excluded. }
\label{fig_deltaf_pdf}
\end{figure}

In Figure~\ref{fig_deltaf}, we map the fraction of the disklike stars
that must be dynamically hot to simultaneously fit the LF and
kinematics (the ``kicked-up fraction''). 
Overall, the kicked-up fraction decreases with radius, as predicted by
\citet{mcc12,pur10}. The kicked-up fraction tends to be higher than the
$1\%$ level predicted by \citet{pur10} for a 1:10 merger event,
though the fraction in each individual subregion is consistent with
$1\%$ at the $1-2\sigma$ level. The fraction of halo stars that come
from the disk (a different quantity from the kicked-up fraction)
varies from $1-30\%$ --- identical to the range found in
the Aquarius halos by \citet{tis13}. Combining all subregions, in
Figure~\ref{fig_deltaf_pdf} we show that the distribution in the overall
kicked-up fraction is $5.2\pm2.1\%$, where the spread in the
distribution comes from both 
the subregion-to-subregion variation and the uncertainty on the mean
value.  The figure also shows that the kicked-up fraction is
greater than $0.6\%$ with $95\%$ confidence. There are two subregions
with outlying kicked-up fractions: $\rm NE1_3$ (with a very high
kicked-up fraction) and $\rm NE2_1$ (where the disklike fraction appears
{\em smaller} than the dynamically cold fraction). 
Subregion $\rm NE1_3$ has a very high kicked-up fraction of disk
stars: $0.31\pm 0.09$. It happens to be at the inner edge of the distribution
of possible locations of the remnant core of the GSS progenitor
\citep{far13}. The broad velocity dispersion in this region could be a
recently disturbed portion of the disk, or be biased by the core itself. 
Subregion $\rm NE2_1$ sits on top of both the dusty 10 kpc ring and the GSS,
so many stars had to be excluded from the
disk fraction measurement. While exclusion of stars in dusty regions
does not appear to bias the velocity distribution, exclusion of
possible GSS debris is much more likely to affect spheroid stars than
disk members. 

Our decomposition is of course limited by our choice of model, and
there are many plausible parameterizations of the bulge, disk and halo
SB profiles. For completeness, we consider the possibility that a
modified SB model could eliminate the need for a kicked-up disk
component. We attempt to fit the data with four different modified SB
models, but none result in an acceptable fit to all three data sets. 

First, we try fitting a model with a second exponential disk component with its
own central intensity, ellipticity, scale length, and LF
parameters. However, the most probable models in the posterior
distribution are those with a single disk component. 

We also try using a model with a single disk whose exponential scale length is
allowed to change at a break radius $R_{\rm break}$ (that is, allowing
for a Freeman type II or III profile). However, the likelihood values
show no preference for a disk with a break. 

Bulgeless disk galaxies can have SB profiles that deviate from pure
exponential. We try fitting a model with a disk whose S\'ersic index
is a free parameter. The resulting PDF of the S\'ersic index has a
mean of $0.94\pm 0.4$ --- that is, the disk is slightly more cored
than a pure exponential. This decomposition does not reduce the
required kicked-up fraction. 

The BHB profile can be well fit by a halo component alone if the halo
core radius is $\sim 3$ kpc \citep{wil12}. We try fixing $R_h = 3.0$
kpc, but this degrades the fit to the  kinematics at large radii
(regions NE2 and NE3) with minimal to no improvement at smaller radii.

The central region of M31 is complex, with multiple
components including a classical bulge that dominates the SB
within 0.2 kpc on the major axis, boxy bulge that dominates within 2.7 kpc, and bar that
extends to at least 4.5 kpc \citep{bea07}. We allow for the
possibility of multiple spheroidal components by fitting a model with
two S\'ersic bulges with unique central intensities, ellipticities,
scale lengths, S\'ersic indices, and LFs. However,
the fit to the kinematics is not improved. 

Finally, we try relaxing the assumption of a S\'ersic bulge profile. The
kinematics would be better fit if the bulge were more
extended --- if it had a shallower slope in the outer regions. However,
the outer slope and inner slopes of a S\'ersic profile are determined by the same
parameter $n$, which in our case is completely constrained by the
inner $\sim 0.7~\rm kpc$ (on the minor axis) where the bulge dominates
the SB. We consider a modified S\'ersic profile, whose inner and
outer slopes are allowed to differ but whose shape reduces to a
S\'ersic when the slopes are the same. Even with this added
flexibility, the data prefer models with near-S\'ersic bulge profiles: 
small bulges nearly identical to those in Figure~\ref{fig_sbp_sersic}.

It is also possible that the excess dynamically hot population belongs
to the virialized remnants of tidal debris. Dynamically cold substructure
has been seen in M31's halo and, to a smaller extent, in the portion
covered by the PHAT survey; it is not unreasonable to suppose that the
remnants from older satellite encounters contribute stars to the
central portion of the galaxy. However, note that the debris would
have to be old enough to
have virialized (because it is dynamically hot), but have a
significant fraction of young or intermediate-age stars (to have
enough AGB stars to fill out the LF brightward of the TRGB). Analysis
of cosmological simulations such as ERIS \citep{gue11} will provide
insight into the relative contributions of young, virialized tidal
debris and kicked-up disk stars in the inner halos of large spiral
galaxies. 


\subsection{Relationship between bulge and halo} 

The location of the transition between M31's bulge and halo has long
been unclear. From images and SB decompositions \citep[e.g.,][]{cou11}
the bulge appears to be relatively small, with a S\'ersic index of
around $2$ and an effective radius of around $1$ kpc. However,
resolved stellar population studies have raised the possibility that the bulge may
be much bigger.  Deep optical HST CMDs of a minor-axis field at a
projected radius of about 15 kpc ($\sim 12$ disk scale radii on the
minor axis) revealed
a broad spread in [Fe/H] ($\sim -1.5 - 0$)
and age ($\sim 5-13$ Gyr) \citep{bro03, bro06}, more similar to the
MW's bulge than to its old, metal-poor inner halo.  It appeared that
either M31 has a large bulge or else its halo has had a much different
formation history than the MW's. 

In \citet{dor12}, we showed that a significant fraction ($\sim
10-20\%$) of the stars in our kinematical sample belong to a
dynamically hot population --- presumably either the outer reaches of a
centrally concentrated bulge or the inner portion of an extended
halo. With kinematics alone, we were unable to distinguish between
these two scenarios, but we are now in a position to show that
the vast majority of these stars are associated with the halo. 

\begin{figure}[h]
\scalebox{0.5}{\includegraphics[trim=40 0 0 30, clip =
  true]{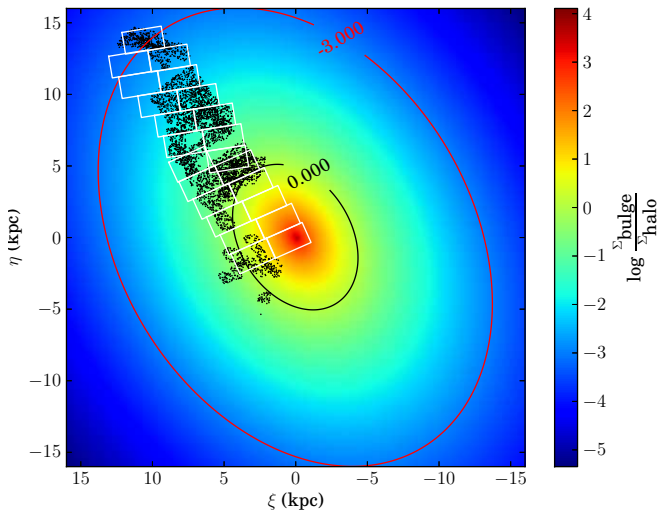}}
\centering
\caption{Map of the relative contribution of the bulge to the spheroid
  SB. The bulge dominates over the halo in the central 5.5 kpc, but
  its surface brightness quickly falls off at larger radii. The spectroscopic sample (black
  dots) falls almost entirely in the region where the spheroid light
  is dominated by the halo; the dynamically hot stars are more likely
  to be associated with the halo than the bulge.}
\label{fig_contour}
\end{figure}

Our decomposition indicates that the SB and LF profiles of the
galaxy are much better fit by a small ($R_b=0.77\pm0.03$ kpc) bulge. 
Figure~\ref{fig_contour} maps the fraction of spheroid SB due to the
bulge. This fraction falls below 0.5 --- that is, the halo dominates
the spheroid SB --- exterior to $R_{\rm proj}\sim 5.5$ kpc on the major
axis. Nearly all of the SPLASH survey region falls in this
halo-dominated region. As shown in \S\ \ref{ssec_diskf}, some
of these stars may have originated in the disk but have been
dynamically heated to kinematically resemble halo members. 

In \citet{dor12}, we found that the dynamically hot population rotates
with $v/ \sigma \sim 1/3$, significantly more slowly than the typical
spiral galaxy bulge \citep{cap07}. It is now clear that this is not
necessarily a useful comparison, as the dynamically hot population is
not physically associated with the bulge. A more relevant comparison
would be to the ``inner halo'' of the MW, which rotates much
more slowly than our dynamically hot population \citep{car07,car10}. It is
possible that the accreted halo population in M31 resembles in the
inner halo of the MW with almost no rotation, but the portion that
originated in the disk has some residual rotation. 

\begin{figure}[h]
\scalebox{0.6}{\includegraphics[trim=0 0 0 30, clip =
  true]{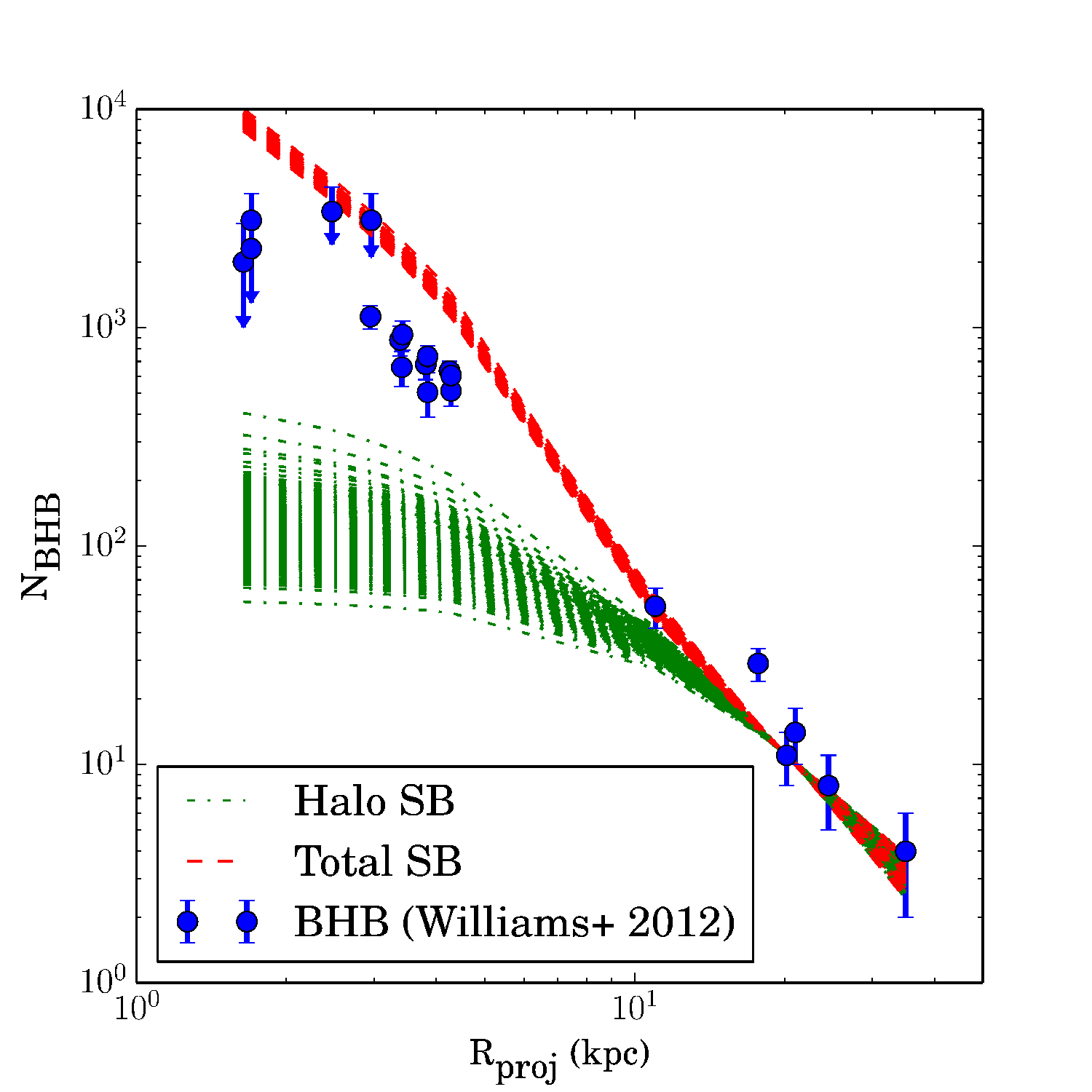}}
\centering
\caption{Comparison between counts of blue horizontal branch (BHB;
  blue dots)
  stars, presumed to trace the metal-poor population \citep{wil12},
  and the most probable halo profile from our decomposition (green
  dot-dashed lines), scaled to
  match the BHB counts at 20 kpc. One line is drawn for each of 256
  points from the posterior probability distribution, so that the
  colored regions approximate the entire allowed (not $1\sigma$)
  region of parameter space. The  power-law halo component cannot explain all of
  the BHB star counts; the bulge and/or the disk must contain a
  significant metal-poor  population. The ratio of BHB counts to SB
  (red lines)
  increases with radius.}
\label{fig_bhb}
\end{figure}

We can learn more about the relationship between the bulge and halo by
comparing their SB profiles to the density profile of BHB stars, a
reliable tracer of metal-poor populations. BHB stars have a mean
metallicity of of [Fe/H]$\sim -1$, with only a small high-metallicity
tail, and so are unlikely to have formed in more metal-rich
subcomponents. \citet{wil12} combined star counts of CMD-selected BHB
stars from the first two years of the PHAT survey. They showed that
the density profile of BHB stars increases steeply interior to 10 kpc,
but not as steeply as the density profile of RGB stars. Our results are qualitatively consistent with theirs.
In Figure~\ref{fig_bhb}, we compare our halo SB profile and total SB
profile (combined halo, bulge, and disk) to the BHB profile of
\citet{wil12}, scaling our SB profiles
to match the BHB counts at $R = 20~\rm kpc$. The BHB/SB ratio (which
should trace the BHB/RGB ratio) increases with radius, as found in
\citet{wil12}. 

However, our halo core radius is large enough that the
halo alone cannot account for the BHB density interior to $10$
kpc. The disk (and possibly the bulge) must supplement the metal-poor
population of the halo. \citet{wil12} argued that it was unlikely that
a significant portion of the BHB stars belonged to the disk, in part
because the kinematically-derived disk fraction in a field at $R=9$
kpc on the minor axis -- where there were many BHB stars -- was no
more than $10\%$. However, we now know two reasons that this disk
fraction may have been an underestimate. First, the SPLASH target selection
strategy preferentially chooses spheroid stars over disk
stars, underestimating the disk fraction by a factor of $1.6$ on the minor
axis. Second, some fraction of the true disk stars may have been
kicked up so that a kinematical decomposition grouped them with
the halo. The metallicity distribution of the disk then is likely quite broad:
some stars belong to a metal-poor BHB population, while the faint
break magnitude of the disk LF indicates a contribution from stars with
solar or supersolar metallicities. 


\subsection{Radial Gradient in the Disk LF}\label{sec_lfgrad}
As shown in Table~\ref{tab_params}, Figure~\ref{fig_dfa} and
Figure~\ref{fig_dfb}, the radial gradient
parameters in the disk LF are exclusively nonzero. 

\begin{figure}[h!] 
\scalebox{1}{\includegraphics[trim=20 185 100 65, clip =
  true]{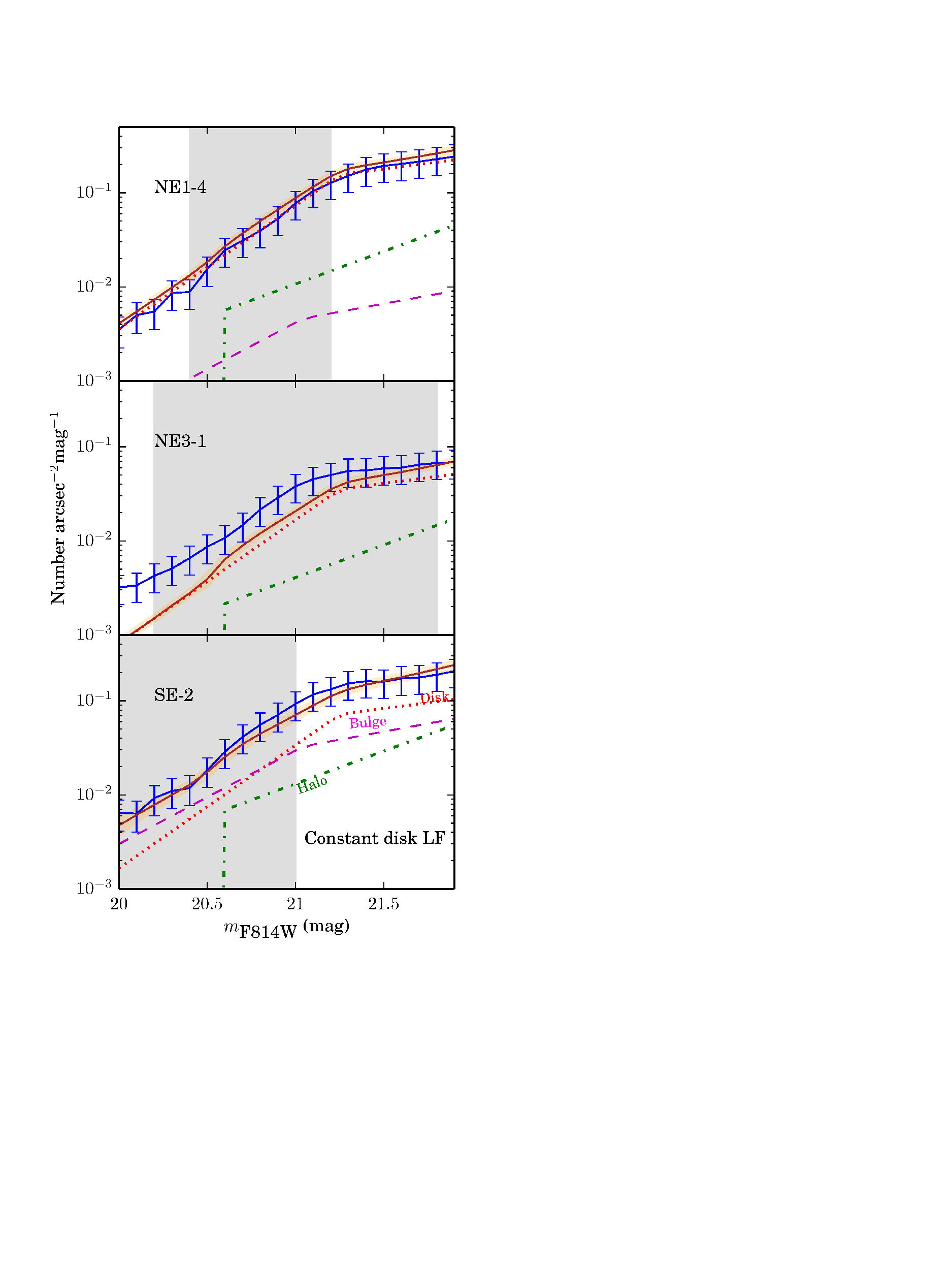}}
\centering
\caption{Same as Figure~\ref{fig_lferr}, assuming a constant disk LF
  shape. The blue error bars, the quadrature sum between the Poisson
  uncertainties and the uncertainty parameter, have been significantly
inflated relative to those in Figure~\ref{fig_lferr} in an attempt to
reduce the tension between the data and this model. A radial gradient
in the disk LF does a much better job of fitting the LF data without
affecting the fits to the SB or kinematical data.}
\label{fig_lferr_nograd}
\end{figure}

To show that a radial gradient is required to simultaneously fit all
three data sets, we run a
test decomposition with a {\em constant} disk LF. The resulting
fit to the LF is quite poor (median reduced
$\chi^2=70$ including only observational
uncertainties). Figure~\ref{fig_lferr_nograd} shows the fits in a
representative sample of three subregions.
Most noticeably, the model predicts a number density that is too high
at small radii and too low at large radii. The uncertainty
parameter on the LF is driven high in an attempt to resolve
the tension between the model and data, inflating the effective
uncertainty beyond the Poisson uncertainty. In contrast, when we allow for a
radial gradient, the LF uncertainty parameter can be very small, and
the structural parameters and the fit to the kinematics are unaffected
(see Figure~\ref{fig_lferr}). 


It is not surprising that there is a radial gradient in the LF of
M31's stellar disk. Such a variation could be caused by either an
age or metallicity gradient, either of which is plausible: HII region
abundance estimates suggest that there may be a small metallicity
gradient within M31's (gas) disk \citep[e.g.,][]{san12, zur12}, while
inside-out disk formation and/or radial migration could induce an age
gradient. 

As discussed earlier in the paper, the break magnitude of the TRGB can
be used as a proxy for metallicity: for populations with
$\textrm{[Fe/H]}\gtrsim -1$,
the TRGB is much more sensitive to metallicity than age, becoming
fainter as metallicity increases. The shapes of the LFs in
Figure~\ref{fig_normlf} illustrate the trend from higher metallicity
at low radius (red curves) to  lower metallicity at high radius
(orange curves). The same
trend -- metal abundance that decreases with radius -- seen in M31
disk PNe \citep{san12, kwi11} and HII regions \citep{san12,
  zur12}. At large radii, the average metallicity of the disk is
similar to that of the bulge, but still higher than that of the halo. 

Note that this gradient exists in the old, smooth stellar disk; it
does not reflect (and thus is not biased by) the sharp changes in
stellar population found in the narrow star-forming rings. Recall
that we excluded from the PHAT
LF every star that fell within a $25\times25~{\rm pc}^2$ pixel with a
reddening $A_v > 1.0$; as shown in the middle panel of
Figure~\ref{fig_datamap}, this cut effectively removes stars in the
star-forming rings. In addition, we have cut out the few remaining
stars blue enough to still be on the young, massive main sequence. 



\section{Summary}\label{sec_summary}
We have presented the first structural decomposition of a large spiral
galaxy using simultaneous SB, LF, and kinematical constraints.  We have
used Bayesian inference to find the probability distribution functions
of 32 parameters describing the surface brightness profile and
LF of each structural subcomponent. We have found: 

\begin{enumerate}
\item The structural parameters we measure are consistent with
  previous measurements. On average, the old stellar disk is more
  highly inclined and its
  major axis PA is larger than measured from isophotal SB fitting. 
\item A decomposition including a S\'ersic bulge, power-law halo,
  exponential disk and constant shapes to the bulge, halo, and disk
  LFs cannot simultaneously well fit the SB, LF, and kinematical
  data. The model poorly predicts the number density of bright stars
 and overestimates the disk fraction in most subregions. 
\item The high kinematically-derived spheroid fractions can be explained if some
  spatially dependent fraction (between $1\%$ and $30\%$) of the
  dynamically hot component is comprised of ``kicked-up'' disk
  stars. This kicked-up population has a disklike SB profile and
  disklike LF, but spheroidlike kinematics ($\sigma_v\sim 150~\rm
  km~s^{-1}$). 
\item In the I band, the halo is brighter than the bulge exterior to
  $R=5.5~\rm kpc$ on the major axis. 
\item Comparison to BHB stars, a tracer of the metal-poor population,
  indicates that the disk metallicity distribution has a
  low-metallicity tail.
\item A SB decomposition including a radially varying disk LF
  improves the fit to the observed LF and does not affect the structural
  parameters.
\end{enumerate}

\begin{deluxetable*}{lllcccccccc}
\tabletypesize{\scriptsize}
\tablecaption{Model Parameters}

\tablewidth{0 pt}
\tablehead{
\colhead{Parameter}&
\colhead{Description}&
\colhead{Units}&
\colhead{Allowed Range\tablenotemark{a}}&
\colhead{Results}&
}
\startdata
$\mu_d$ & Disk central SB & mag & $0<\mu_d<\infty$ & $18.901^{+0.032}_{-0.031}$\\
\vspace{2mm}
$R_d$ & Disk scale length & kpc & $0<R_d<\infty$ & $5.76^{+0.101}_{-0.113}$\\
\vspace{2mm}
$\epsilon_d$ & Disk ellipticity & 1 & $0<\epsilon_d<1$ & $0.725^{+0.005}_{-0.005}$\\
\vspace{2mm}
$\mu_b$ & Bulge SB at $R_e$ & mag & $0<\mu_b<\infty$ & $17.849^{+0.066}_{-0.066}$\\
\vspace{2mm}
$R_b$ & Bulge half-light radius & kpc & $0<R_b<\infty$ & $0.778^{+0.03}_{-0.028}$\\
\vspace{2mm}
$n_b$ & Bulge Sersic index & 1 & $0<n_b<10$ & $1.917^{+0.082}_{-0.081}$\\
\vspace{2mm}
$\epsilon_b$ & Bulge ellipticity & 1 & $0<\epsilon_b<1$ & $0.277^{+0.011}_{-0.011}$\\
\vspace{2mm}
$\mu_h$ & Halo central intensity & mag & $0<\mu_h<\infty$ & $24.18^{+0.295}_{-0.328}$\\
\vspace{2mm}
$R_h$ & Halo scale radius & kpc & $0<R_h<\infty$ & $10.631^{+2.459}_{-2.034}$\\
\vspace{2mm}
$\epsilon_h$ & Halo ellipticity & 1 & $0<\epsilon_h<1$ & $0.215^{+0.109}_{-0.12}$\\
\vspace{2mm}
$\alpha_h$ & Halo power-law slope & 1 & $0<\alpha_h<\infty$ & $2.508^{+0.232}_{-0.199}$\\
\vspace{2mm}
$\textrm{PA}$ & Major axis PA & degrees & $0<\textrm{PA}<180$ & $6.632^{+0.459}_{-0.511}$\\
\vspace{2mm}
$y_{d0}$ & Disk \# density/SB, $R_{\textrm{disk}}$ = 1 kpc & $10^5\textrm{N}/\textrm{arcsec}^2$/mag & $0<y_{d0}<\infty$ & $1.93^{+4.00}_{-3.90}$\\
\vspace{2mm}
$p_{d0}$ & Disk LF bright-end slope, $R_{\textrm{disk}}$ = 1 kpc & log(N)/mag & $0<p_{d0}<\infty$ & $4.519^{+0.313}_{-0.267}$\\
\vspace{2mm}
$q_{d0}$ & Disk LF faint-end slope, $R_{\textrm{disk}}$ = 1 kpc & log(N)/mag & $0<q_{d0}<\infty$ & $2.012^{+0.826}_{-0.843}$\\
\vspace{2mm}
$l_{d0}$ & Disk LF break magnitude, $R_{\textrm{disk}}$ = 1 kpc & mag & $20<l_{d0}<22$ & $21.782^{+0.146}_{-0.209}$\\
\vspace{2mm}
$\delta y_d$ & Radial gradient in disk \# density/SB & N/$\textrm{arcsec}^2$/mag/ln(kpc) & $-\infty<\delta y_d<\infty$ & $0.001^{+0.0}_{-0.0}$\\
\vspace{2mm}
$\delta p_d$ & Radial gradient in disk LF bright-end slope & log(N)/mag/ln(kpc) & $-\infty<\delta p_d<\infty$ & $-0.21^{+0.083}_{-0.06}$\\
\vspace{2mm}
$\delta q_d$ & Radial gradient in disk LF faint-end slope & log(N)/mag/ln(kpc) & $-\infty<\delta q_d<\infty$ & $-0.777^{+0.111}_{-0.134}$\\
\vspace{2mm}
$\delta l_d$ & Radial gradient in disk LF break magnitude & mag/ln(kpc) & $-\infty<\delta l_d<\infty$ & $-0.672^{+0.321}_{-0.308}$\\
\vspace{2mm}
$y_b$ & Bulge \# density/SB & N/$\textrm{arcsec}^2$/mag & $0<y_b<\infty$ & $0.007^{+0.002}_{-0.001}$\\
\vspace{2mm}
$p_b$ & Bulge LF bright-end slope & log(N)/mag & $0<p_b<\infty$ & $2.494^{+0.405}_{-0.38}$\\
\vspace{2mm}
$q_b$ & Bulge LF faint-end slope & log(N)/mag & $0<q_b<\infty$ & $0.424^{+0.434}_{-0.307}$\\
\vspace{2mm}
$l_b$ & Bulge LF break magnitude & mag & $20<l_b<22$ & $21.125^{+0.144}_{-0.189}$\\
\vspace{2mm}
$y_h$ & Halo \# density/SB & N/$\textrm{arcsec}^2$/mag & $0<y_h<\infty$ & $0.014^{+0.004}_{-0.003}$\\
\vspace{2mm}
$q_h$ & Halo LF faint-end slope & log(N)/mag & $0<q_h<\infty$ & $1.592^{+0.301}_{-0.351}$\\
\vspace{2mm}
$l_h$ & Halo LF break magnitude & mag & $20.3<l_h<22$ & $20.805^{+0.046}_{-0.052}$\\
\vspace{2mm}
$\epsilon_{\textrm{LF}}$ & LF uncertainty parameter & 1 & $0<\epsilon_{\textrm{LF}}<\infty$ & $0.129^{+0.008}_{-0.007}$\\
\vspace{2mm}
$\epsilon_{\textrm{Choi}}$ & Uncertainty on Choi SB data & 1  & $0<\epsilon_{\textrm{Choi}}<2$ & $0.086^{+0.005}_{-0.005}$\\
\vspace{2mm}
$\epsilon_{\textrm{Gilbert}}$ & Uncertainty on Gilbert SB data & 1 & $0<\epsilon_{\textrm{Gilbert}}<2$ & $0.815^{+0.155}_{-0.124}$\\
\vspace{2mm}
$\epsilon_{\textrm{PvdB}}$ & Uncertainty on PvdB SB data & 1 & $0<\epsilon_{\textrm{PvdB}}<2$ & $0.577^{+0.7}_{-0.333}$\\
\vspace{2mm}
$\epsilon_{\textrm{Irwin}}$ & Uncertainty on Irwin SB data & 1 & $0<\epsilon_{\textrm{Irwin}}<2$ & $0.086^{+0.005}_{-0.005}$\\
\vspace{2mm}

\enddata
\tablenotetext{a}{We used a flat prior within the range indicated.} 
\label{tab_params}

\end{deluxetable*}

\section{Acknowledgments}

We thank Stephane Courteau, Alis Deason, Dimitrios
Gouliermis, James Guillochon, Katherine Hamren, Antonela Monachesi,
and Annalisa Pillepich for helpful discussions and feedback. PG and CD
acknowledge NSF grants AST-0607852 and AST-1010039 and NASA grant
HST-GO-12055. KG acknowledges Hubble Fellowship grant 51273.01 awarded
by the Space Telescope Science Institute, which is operated by the
Association of Universities for Research in Astronomy, Inc., for NASA,
under contract NAS 5-26555. Additionally, CD was supported by a NSF Graduate
Research Fellowship. 

We acknowledge the very significant
cultural role and reverence that the summit of Mauna Kea has always
had within the indigenous Hawaiian community. We are most fortunate to
have the opportunity to conduct observations from this mountain. 


\clearpage

\appendix

\section{Spectroscopic Data} \label{appendix_specdata}
The details of the spectroscopic target selection and observation are given
in \citet{dor12}. However, here in Table~\ref{tab_obs}, we describe
the five additional slitmasks that were observed with Keck/DEIMOS in
November 2011 and therefore were not included in \citet{dor12}. 

All of the targets on these five masks were candidate RGBs identified
on the basis of the optical CMD from the PHAT survey. 

\begin{deluxetable*}{llllrccccc}[h!]
\tabletypesize{\scriptsize}
\tablecaption{Keck/DEIMOS Multiobject Slitmask Exposures from Fall 2011}
\tablewidth{0 pt}
\tablehead{
\colhead{Mask}&
\colhead{Observation}&
\colhead{$\alpha$ [J2000]}&
\colhead{$\delta$ [J2000]}&
\colhead{P.A.}&
\colhead{$t_{\rm{exp}}$}&
\colhead{Seeing}&
\colhead{No. of}&
\colhead{No. of Usable}&
\colhead{No. of Usable}\\
\colhead{Name}&
\colhead{Date (UT)}&
\colhead{(h m s)}&
\colhead{($^\circ$~$'$~$''$)}&
\colhead{($^\circ$)}&
\colhead{(sec)}&
\colhead{FWHM}&
\colhead{Slits}&
\colhead{Target Velocities}&
\colhead{Velocities of}\\
\colhead{}&
\colhead{}&
\colhead{}&
\colhead{}&
\colhead{}&
\colhead{}&
\colhead{}&
\colhead{}&
\colhead{(Success Rate)}&
\colhead{Serendipitously Detected}\\
\colhead{}&
\colhead{}&
\colhead{}&
\colhead{}&
\colhead{}&
\colhead{}&
\colhead{}&
\colhead{}&
\colhead{}&
\colhead{Stars}
}
\startdata
	mctF & 2011 Nov 23 & 00 44 24.00 & $+$41 36 00.0  & $-$30.0
	& 2900 & $0\farcs6$ & 246 & 179 (73\%) & ~47\\
        mctG &  2011 Nov 23 & 00 45 53.03 & $+$41 42 05.1 & $+$25.0
        & 2900 & $0\farcs5$ & 259 & 207 (80\%) & ~24\\        
        mctJ & 2011 Nov 24 & 00 45 10.80 & $+$41 55 48.0 & $+$35.0
	& 3600  & $0\farcs6$ & 253 & 182 (72\%) & ~18\\ 
        mctK&  2011 Nov 24 & 00 46 46.85 & $+$42 13 35.3 & $+$45.0
	& 3400 & $0\farcs8$  & 270 & ~208 (78\%) & ~10\\	
     	mctL &  2011 Nov 23 & 00 46 19.97 & $+$42 14 05.2& $-$65.0
	& 3680  & $1\farcs0$ & 257 & 182 (71\%) & ~~7 \\
\tableline\tableline \vspace{1 mm}
        \\
	Total: &&&&&&& 1285~ & 958 (75\%)~ & 106
\enddata

\label{tab_obs}
\end{deluxetable*}

\section{MCMC Sampler} \label{appendix_mcmc}
To estimate the marginalized posterior probability function of the model
parameters, we draw samples from the distribution $P$
(Equation~\ref{eq:likelihood}) using a Markov
chain Monte Carlo sampler \citep[MCMC;][]{bis03,gel03,mac03,pre07}. MCMC
algorithms offer a method of
efficiently drawing unbiased samples from any distribution
that can be evaluated for a given set of parameters. In this case, the
distribution of interest is the posterior probability which is---up to an
unimportant normalization constant---the product
of the likelihood function (Equation~\ref{eq:likelihood}) and the
prior distributions over the parameters (which we choose to be flat
for all parameters). Drawing samples from this
distribution is equivalent to drawing a representative sampling of physical
models that are consistent with the data and the uncertainties.

The most popular class
of MCMC algorithms are based on the Metropolis--Hastings (M--H) method
\citep[e.g.,][]{mac03}. For this project we use a more efficient ensemble
sampler called \emcee\ that
takes advantage of an affine-invariant proposal distribution in order to
sample efficiently even in arbitrarily covariant parameter spaces
\citep{goo10, for12}. In many cases, \emcee\
requires many fewer computations than required by a standard M--H algorithm
to draw the same number of independent samples from a distribution. The
advantage comes from the fact that \emcee\ uses an affine-invariant
algorithm, meaning that it performs equally well regardless of
the covariance between parameters. If a M--H chain is to perform well, the
proposal distribution---which generally has $\sim D^2$ free
parameters, where D is the number of dimensions in the problem---must
be tuned to match the covariances in the target density using a
computationally expensive ``burn-in'' phase. In contrast, \emcee\ has only 2
free parameters and the proposal density adaptively fits the target density
without any fine tuning. This is achieved by simultaneously evolving
$K$ 
coupled MCMC chains or ``walkers''. Each walker produces unbiased samples from
the target distribution but the proposal distribution for each walker is
determined by the current positions of all the other walkers. For more
information about this algorithm and the implementation details, see the
discussion in \citet{for12}.

For this project, we use 256 walkers and keep the \emcee\ proposal
scale---called $a$ in \citet{for12} ---at the default value of
2. Each MCMC run is composed of $\sim 10,000$ ``burn-in'' steps for each
walker and then $100$ production
steps. The results given in the paper are based on the $256 \times 20
= 51,200$ samples produced after burn-in. In particular, the final column in
Table~\ref{tab_params} reports the sample mean and standard deviation
of the parameters in the production chain.

\section{Parameter Distributions}
\begin{figure*}[h] 
\scalebox{1}{\includegraphics[trim=30 60 10 60, clip =
  true]{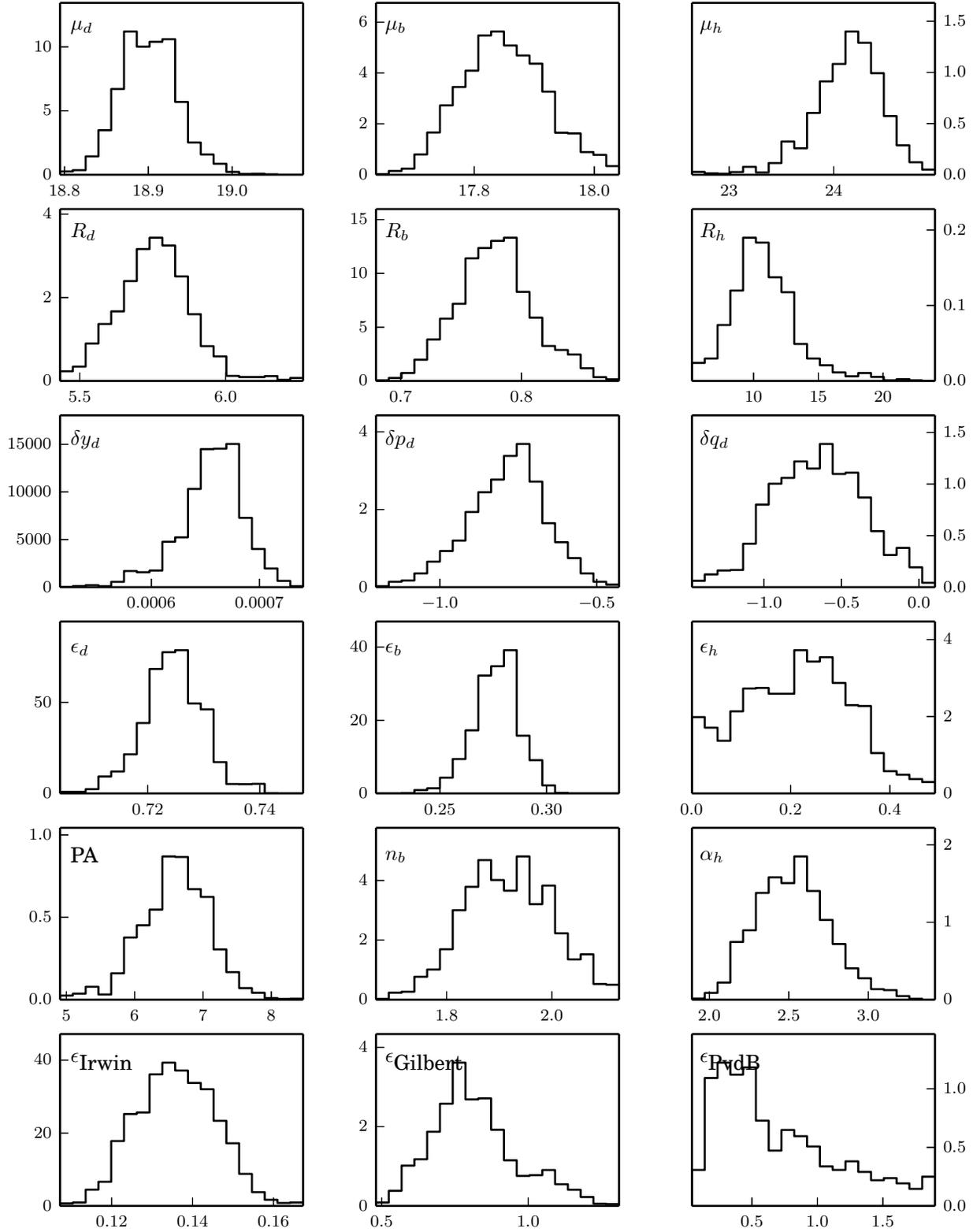}}
\centering
\caption{Posterior probability distributions of each parameter, as
  drawn from the last 20 steps of the MCMC chain. Units
  are given in Table~\ref{tab_params}. With the exception of the third
  row, parameters describing the disk, bulge, and halo are shown in the
  left, middle, and right columns, respectively. }
\label{fig_dfa}
\end{figure*}

\begin{figure*}[h] 
\scalebox{1}{\includegraphics[trim=30 50 10 60, clip =
  true]{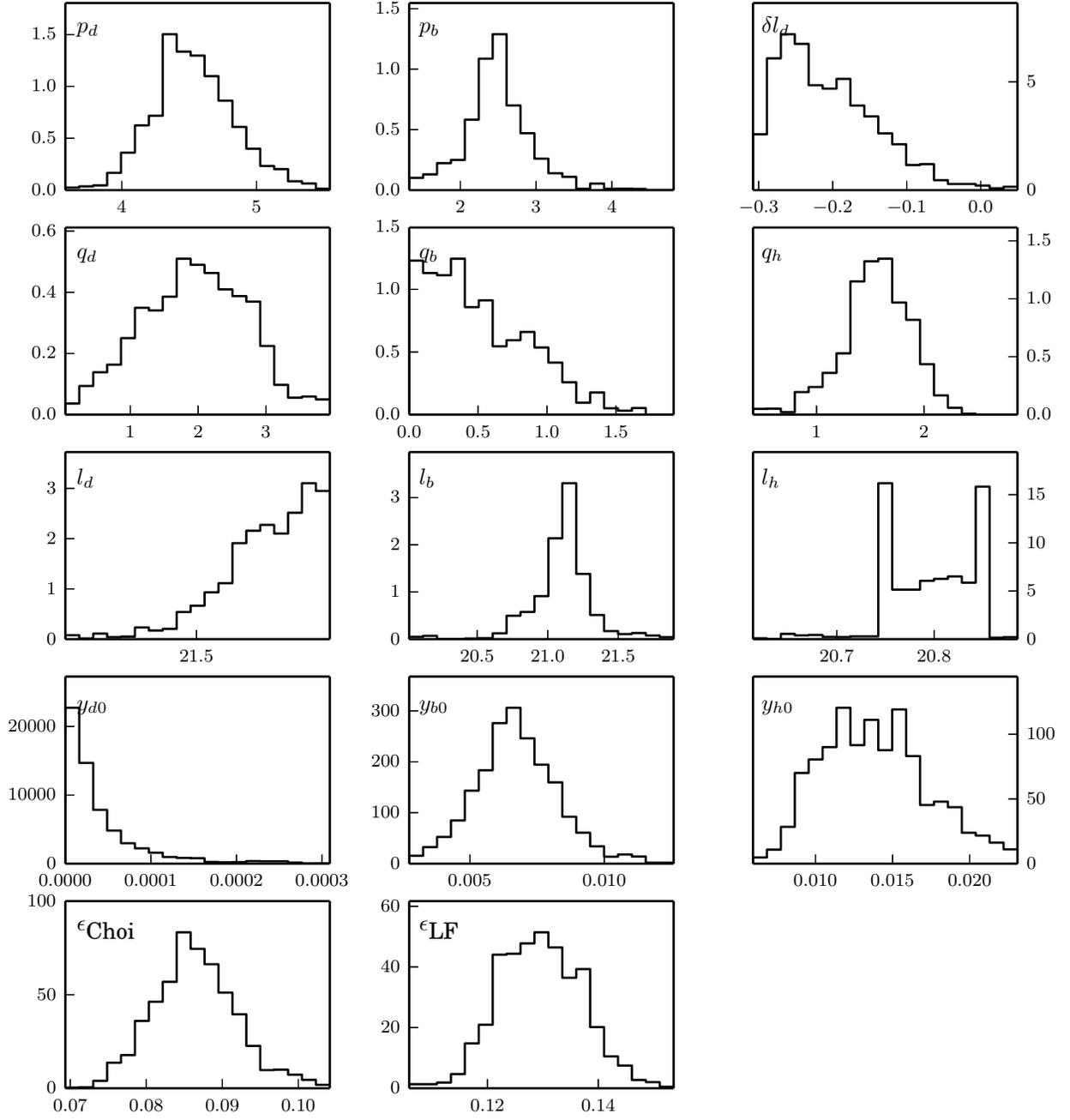}}
\centering
\caption{Figure~\ref{fig_dfa}, continued.}
\label{fig_dfb}
\end{figure*}

Figures~\ref{fig_dfa} and \ref{fig_dfb} show the 1D posterior probability
distributions of all model parameters using samples from the last
$20$ steps of the MCMC chain. Each distribution is normalized to an
area of 1 for display purposes. Descriptions of parameters are given
in Table~\ref{tab_params}.

\end{document}